\documentclass[prx,twocolumn,floatfix,superscriptaddress,longbibliography]{revtex4-2}

\usepackage[dvips]{graphicx}
\usepackage{amsmath,amssymb,amsthm,mathrsfs,amsfonts,dsfont,mathtools,physics}
\usepackage{epsfig}
\usepackage{braket}
\usepackage[colorlinks=true,citecolor=blue,linkcolor=magenta]{hyperref}
\usepackage{bm}
\usepackage{enumerate}
\usepackage{color}
\usepackage{graphicx}
\usepackage{tikz}
\usetikzlibrary{quantikz2}
\usepackage{appendix}

\usepackage{enumitem}
\usepackage[capitalize]{cleveref}
\usepackage[normalem]{ulem}

\usepackage{amsthm}
\newtheorem{result}{Result}

\newtheorem{statement}{Statement}

\Crefname{theorem}{Theorem}{Theorems}
\theoremstyle{remark}

\renewcommand{\tr}{\mathrm{Tr}}

\renewcommand{\var}{\mathrm{Var}}

\newcommand{\qmaddress}{\affiliation{Quantum Motion, 9 Sterling Way, London N7 9HJ, United Kingdom}}
\newcommand{\oxaddress}{\affiliation{Department of Materials, University of Oxford, Parks Road, Oxford OX1 3PH, United Kingdom}}
\newcommand{\lanladdress}{\affiliation{Theoretical Division, Los Alamos National Laboratory, Los Alamos, New Mexico 87545, USA}}

\begin{document}

\title{Direct Estimation of the Density of States for Fermionic Systems}

\author{Matthew L. Goh}
\email{matthew.goh@quantummotion.tech}
\oxaddress
\lanladdress
\qmaddress

\author{B\'alint Koczor}
\email{koczor@maths.ox.ac.uk}
\affiliation{Mathematical Institute, University of Oxford, Woodstock Road, Oxford OX2 6GG, United Kingdom}
\oxaddress
\qmaddress

\begin{abstract}
Simulating time evolution is one of the most natural applications of quantum
computers and is thus one of the most promising prospects for achieving practical quantum advantage.
Here, we develop quantum algorithms to extract thermodynamic properties
by estimating the density of states (DOS), which is a central object in quantum statistical mechanics.
We introduce several key innovations that significantly improve the practicality and extend
the generality of previous techniques.
First, our approach allows one to estimate the DOS only for a specific subspace of the full Hilbert space.
This is crucial for fermionic systems, since both canonical and grand canonical ensemble thermal equilibrium properties depend on subspaces of fixed number.
Second, in our approach, by time evolving very simple, random initial states, such as randomly chosen computational basis states,
we can exactly recover the DOS on average.
Third, due to circuit-depth limitations, we only reconstruct the DOS up to a convolution with a Gaussian window --
thus all imperfections that shift the energy levels by less than the width of the convolution window will
not significantly affect the estimated DOS.
For these reasons we find the approach is a promising candidate for early quantum advantage as even short-time, noisy dynamics
can yield a semi-quantitative reconstruction of the DOS (convolution with a broad Gaussian window), while early
fault tolerant devices will likely enable higher resolution DOS reconstruction through longer time evolutions.
We demonstrate the practicality of our approach in representative Fermi-Hubbard and spin
models and indeed find that our approach is highly robust against algorithmic errors in the time evolution and against gate noise. We further demonstrate that our approach is compatible with NISQ-friendly variational techniques, introducing and leveraging a new technique for variational time evolution.
\end{abstract}

\maketitle

\section{Introduction}

The primary application area of quantum computers will likely be the simulation of many-body quantum 
systems, such as in quantum field theory~\cite{latticeschwinger}, quantum gravity~\cite{jafferis2022traversable},
as well as materials and chemical systems~\cite{cao2019quantum, mcardle2020quantum,bauer2020quantum, Motta2022}.
For these reasons a broad range of techniques have been developed for simulating the time-evolution under
a given problem Hamiltonian, such as product formula or Trotterisation~\cite{berry2007efficient}, quantum signal processing~\cite{Low2017optimal}, qubitization~\cite{Low2019hamiltonian} or
linear combination of unitaries~\cite{Berry2015}.

The circuit depth of these techniques grows proportionally with the total simulation time
and may thus enable quantum advantage for certain systems, given that 
it is in general exponentially hard to simulate quantum systems using classical techniques.
In practice, however, it still remains highly challenging to achieve quantum advantage for quantum dynamics, since
long time evolutions at scale would require quantum error correction, which requires significantly
more advanced hardware.
For these reasons, a broad range of quantum algorithms have been proposed using shallow quantum circuits
and quantum error mitigation for the purposes of using
early, noisy quantum devices in quantum simulation tasks~\cite{kim2023evidence, error_mitigation,PhysRevX.11.031057,PhysRevX.12.041022,farhi2014quantum,peruzzo2014variational,endoHybridQuantumClassicalAlgorithms2021, cerezoVariationalQuantumAlgorithms2021a, bharti2021noisy,chan2022algorithmic,li2017efficient,PhysRevLett.125.180501}.

Here we develop a range of new and improved quantum algorithms for estimating thermodynamic properties
of quantum systems that commonly apply time evolution to a series of randomly initialised, simple,
starting states. 
In particular, we focus on the density of states (DOS), a central object
that enables the computation
of various thermodynamic properties important for chemical and materials modelling. 
A broad range of classical simulation techniques have been explored for direct estimation of the density of states~\cite{weisse2006kernel,PhysRevE.75.046701,shell2003improved,kong2022density}, however it is known to be a computationally hard problem in general~\cite{brown2011computational}. Due to the exponential classical scaling of quantum simulation problems, quantum computers will likely prove advantageous in computing the DOS. Several prior works have introduced quantum algorithms to directly estimate the DOS, via Monte Carlo methods utilizing time evolution \cite{lu2021algorithms,schuckert2023probing,ghanem2023robust} or by porting classical kernel polynomial methods \cite{weisse2006kernel} to quantum computers \cite{summer2024calculating}. These prior approaches have the commonality of utilizing randomized initial states for Hadamard tests on controlled unitary evolution, and were all demonstrated on simple spin-chain models.

However, since the constituents of matter are fermions, almost all \emph{ab initio} quantum models of genuine utility in applied fields like chemistry and materials science will have a fermionic component, with a Fock space of variable particle number.
Dealing with this is unavoidable when studying fermionic Hamiltonians on quantum computers, since the mapping-dependent encoding of particle number to the qubit register must be accounted for by the quantum algorithm. For example, in the Jordan-Wigner mapping, the particle number is directly encoded in the Hamming weight of computational basis states; various other fermion-to-qubit mappings with different advantages and disadvantages have been developed, where the particle number may not be as easily identifiable.
Furthermore, equilibrium thermal properties in both the canonical and grand canonical ensembles directly depend on the DOS in fixed-number subspaces (i.e. the subspace DOS only includes energies of eigenstates within this fixed-number subspace - see Equations \ref{eqn:canonical_partition_function}, \ref{eqn:subspace_DOS_definition} and \ref{eqn:grand_canonical_partition_function}), so efficiently computing it is a crucial step in leveraging quantum computers for practical problems in chemistry and materials modelling.

These prior approaches \cite{lu2021algorithms,schuckert2023probing,ghanem2023robust,summer2024calculating} were only demonstrated on simple spin models and not generalized to fermionic problems, limiting their utility for high-value practical problems. For fermionic problems, it is often prudent to indirectly compute the DOS via Green's functions. For systems with relatively weak correlations, one can therefore draw upon the rich background of Green's function methods from the condensed matter community, particularly perturbative calculations with Feynman diagrams~\cite{abrikosov1963methods,Keldysh:1964ud,hedin1965new,fetter1971quantum,mahan2013many}. However, the frontier of study in many-body quantum systems is concerned with strongly correlated systems well outside of the perturbative regime. A variety of quantum algorithms for the computation of fermionic Green's functions exist~\cite{bauer2016hybrid,rungger2019dynamical,endo2020calculation,jamet2021krylov,jamet2023anderson}, but these methods typically require explicit preparation of ground and/or excited states of the Hamiltonian, which is a highly non-trivial problem in general.

In this work, we bridge this gap, combining the simplicity of random-state Hadamard-test methods with the utility of fermionic systems. Our approach brings several significant innovations. First, our approach allows us to compute the DOS on \emph{any} subspace of the full Hilbert space for which a basis can be prepared, which crucially includes fermionic subspaces of fixed number.

Second, our approach applies time evolution to very simple random initial states. In particular, we prove that unitary 1-design random circuits allow us to exactly recover the DOS on average, and thus circuits as simple as single-qubit bit-flips can be used. This represents a major improvement compared to prior techniques using more demanding random initializations, such as 2- and 3-design circuits \cite{PhysRevA.93.062306,lu2021algorithms,schuckert2023probing,ghanem2023robust,summer2024calculating}. While our circuits in \cref{fig:circ} appear similar to statistical phase estimation circuits due
to the use of controlled time evolution, we note that our circuits are guaranteed to reconstruct the
full density of states up to a broadening of lines and, in contrast to phase estimation, do not
require one to supply a ``good enough'' initial state -- which would be exponentially hard in general.

Third, we use a simple windowed Fourier transform to estimate the DOS up to 
a broadening of lines which allows us to make clear tradeoffs between resolution in the DOS
vs the total time-evolution length -- given algorithmic errors in the time evolution, as well
as gate noise in the quantum circuits limit circuit depths and total time evolution lengths,
estimating relatively low-resolution DOS can be extremely robust to noise sources.

We numerically simulate a range of practical examples including simple spin systems and Fermi-Hubbard
models and analyse the effect of both algorithmic errors in the time evolution as well as
gate noise. We find the approach is highly robust against various sources of errors as long as 
the required resolution is relatively low (convolution window width is relatively broad).
We also investigate a range of random-state initialisation techniques and find that very simple 
random single-qubit rotations are very practical. Our approach is thus a powerful and simple 
candidate for early practical advantage that uses quantum computers for relatively short time evolutions
(but still longer than accessible with classical techniques) and is applicable to fermionic quantum systems.

In the rest of this introduction we introduce basic notions in the context of quantum thermodynamics and
introduce the DOS both for fixed and for variable particle numbers. In \cref{sec:results} we then
introduce our main results using both deterministic mixed initial state and using
random pure initial states.  In \cref{sec:numerics} we perform a broad range of numerical
experiments to verify the practical utility of the present approach, finding that our approach is extremely resilient to practical concerns such as shot noise, algorithmic error, and hardware noise, as well as introducing a new method for variational time evolution and utilizing it to demonstrate a NISQ-friendly variant of our techniques. Finally, we draw conclusions from our work in \cref{sec:conclusion}.

\subsection{Quantum systems at finite temperature}
Our aim in this work is to estimate thermal properties of quantum systems at finite temperature
by analysing the time evolution of either randomly initialised pure quantum states or to deterministic mixed states.
Before stating our main results, we here briefly review the relevant definitions and quantities.

\subsubsection{Physical systems of constant particle number}\label{sec:constant_particle_numb}
Let us first consider a $d$-dimensional quantum system
with a Hamiltonian $\hat{H}$ defined only for a fixed number of particles, such as a spin system. 
A key quantity required to determine thermal properties in the canonical ensemble
(i.e., equilibrium with a thermal reservoir at inverse temperature $\beta$)
is the \emph{partition function} which is defined as
\begin{align} \nonumber
Z(\beta)& := \Tr[e^{-\beta \hat{H}}]\\
    &=\int dE e^{-\beta E}g(E).
    \label{eqn:trivial_canonical_Z}
\end{align}
Here $g(E)$ is the \emph{density of states} (DOS) which is defined as a series
of Dirac-deltas at the Hamiltonian eigenenergies as
\begin{equation}
    g(E) :=\sum_{k=1}^d \delta(E-E_k).
    \label{eqn:trivial_DOS}
\end{equation}
The above density of states is thus a central quantity as it allows one to determine $Z(\beta)$,
and therefore any further thermodynamic property,
e.g., the internal energy is obtained as $U(\beta)=-\partial (\log Z)/\partial \beta$.
In the present work we devise new quantum algorithms for estimating $g(E)$
that improve upon and generalise algorithms obtained in related prior
works~\cite{lu2021algorithms,schuckert2023probing,ghanem2023robust,summer2024calculating}.

\subsubsection{Physical systems of variable particle number}\label{sec:variable_particle_numb}
A broad range of physical systems encountered in practice are defined on Hilbert spaces
that simultaneously accommodate multiple different particle numbers.
For such systems we need to go beyond the methods in previous works ~\cite{lu2021algorithms,schuckert2023probing,ghanem2023robust,summer2024calculating}; as we detail now,
 the central object we need to determine is the subspace density of states $g_M(E)$.
For example, fermionic Fock spaces in quantum chemistry and condensed matter physics need to account for multiple, varying particle numbers
-- consequently, naively applying the conventional expression in \cref{eqn:trivial_DOS} would simultaneously contain information about
multiple particle numbers.
However, one typically models systems of fixed particle number and thus focuses only
on the relevant subspace of the Hilbert space that corresponds to the relevant, fixed particle number.

More specifically, when the system's Hamiltonian $\hat{H}$ commutes with a total number operator $N$ as $[\hat{H},N]=0$, then eigenstates $\ket{k,m}$ can be labelled by a quantum number $k \in \mathcal{S}_m$ within a fixed subspace of total particle number $m$ (here $\mathcal{S}_m$ collects all indices that belong to the same particle number).
For example, all computational basis states of Hamming weight $m$ (total number of 1s in. e.g., $\ket{1001\dots}$) form a subspace of total particle number $m$ when encoding fermionic states into qubit states using the Jordan-Wigner encoding.

First, we consider the canonical-ensemble partition function for a fixed particle number which is defined as
\begin{align}
    Z_M(\beta)&\equiv \Tr_M[e^{-\beta \hat{H}}]\nonumber\\
    &= \int dE e^{-\beta E}g_M(E).
\label{eqn:canonical_partition_function}
\end{align}
Indeed, the above is closely related to \cref{eqn:trivial_canonical_Z}
but the trace only includes eigenstates contained in the subspace of particle number $M$
as $\Tr_M[\cdot]=\sum_{k\in \mathcal{S}_M}\bra{k,M}\cdot\ket{k,M}$.
Furthermore, the density of states on this subspace is defined as
\begin{equation}\label{eqn:subspace_DOS_definition}
    g_M(E)\equiv \sum_{k=1}^{k_M}\delta(E-E_{k,M}).
\end{equation}

Second, another application of our approach is to compute thermal properties in the grand canonical ensemble
whereby the particle number is permitted to fluctuate, i.e., 
equilibrium with a thermal and particle reservoir at inverse temperature $\beta$ and chemical potential $\mu$.
The corresponding grand canonical partition function is computed as
\begin{align}
    \mathcal{Z}(\beta,\mu)&\equiv \Tr[e^{-\beta(\hat{H}-\mu N)}]\nonumber\\
    &= \sum_M e^{\beta \mu M}Z_M(\beta),
    \label{eqn:grand_canonical_partition_function}
\end{align}
a weighted sum of canonical partition functions at varying particle number $M$, weighted by powers of the fugacity $z=e^{\beta\mu}$.

\begin{figure}
    \centering
    \includegraphics[width=0.48\textwidth]{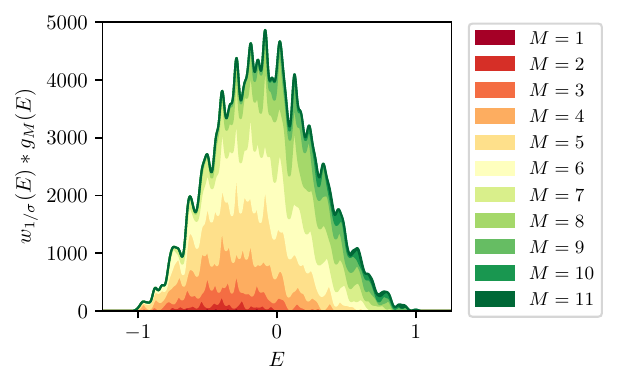}
    \caption{\textbf{Density of states of a Hubbard model on subspaces of fixed number $M$.}
    We cumulatively plot $g_M(E)$ (up to a broadening of lines discussed in \cref{sec:window_width_effects})
    for a $(3\times 2)$ open-boundary grid Hubbard model \eqref{eqn:fermi_hubbard_hamiltonian} with $J=-1$, $U=2$
    for a series of different particle numbers $M$. Each $M$ has a different DOS (different colours) and a different normalization factor
    $|\mathcal{S}|$ which has important implications for computing the partition function $Z_M(\beta)$
    or the grand canonical partition function $\mathcal{Z}(\beta,\mu)$. Here, $M=6$ is the most populous subspace.}
    \label{fig:subspace_spectra}
\end{figure}

In both cases of the canonical and grand canonical ensembles, determining the subspace density of states $g_M(E)$ is
sufficient to access all equilibrium thermal properties. 
Some of the main contributions of the present work are a set of techniques for estimating $g_M(E)$
for fermionic systems on quantum computers;
While the relevant theory is detailed in \cref{sec:dos_with_circuits}, we demonstrate our methods
in a set of numerical experiments in \cref{sec:numerics}
using the Fermi-Hubbard model, which is an
archetypal example of a fermionic many-body Hamiltonian  as
\begin{equation}
    \hat{H} = -J\sum_{\langle \bm{j}, \bm{k} \rangle, \sigma}\left(c^\dagger_{\bm{j}\sigma}c_{\bm{k}\sigma}+c^\dagger_{\bm{k}\sigma}c_{\bm{j}\sigma}\right)+U\sum_{\bm{j}}n_{\bm{j}\uparrow}n_{\bm{j}\downarrow}.
    \label{eqn:fermi_hubbard_hamiltonian}
\end{equation}
Here $c_{\bm{j}\sigma}$ ($c^\dagger_{\bm{j}\sigma}$) is the fermionic annihilation (creation) operator for site $\bm{j}$ and spin $\sigma\in\{\uparrow,\downarrow\}$, $n_{\bm{j}\sigma}=c^\dagger_{\bm{j}\sigma}c_{\bm{j}\sigma}$ is the spin-density operator for spin $\sigma$ on site $\bm{j}$, and $\langle \bm{j}, \bm{k} \rangle$ indicates summation over neighbouring sites. Since $[\hat{H},N]=0$ for total number $N=\sum_{\bm{j}}(n_{\bm{j}\uparrow}+n_{\bm{j}\downarrow})$, its eigenstates have well-defined total number $M$. 
In \cref{fig:subspace_spectra} we illustrate the density of states $g_M(E)$ at a varying $M$ for a
particular instance of the Fermi-Hubbard Hamiltonian (although not actually required for the algorithm
as we detail in \cref{sec:window_width_effects}, for convenient normalization of variables we rescale our
Hamiltonian such that its maximum/minimum eigenvalues are $\pm1$).

\section{Results}
\label{sec:results}
A central object in accessing thermodynamic properties through time evolution will
be the Fourier transform of the DOS, which we now briefly define and relate to the time-evolution operator.
\begin{statement}\label{state:fdos}
	For physical systems with fixed particle number, the Fourier transform of the density of states (FDOS) $G(t) := \mathcal{F}[g(E)](t) $ can be computed as 
	the trace of the time-evolution operator as
	\begin{equation}\label{eqn:FDOS_definition_trivial}
		G(t) = \frac{1}{\sqrt{2\pi}}\Tr[e^{-i\hat{H}t}].
	\end{equation}
\end{statement}
Please refer to \cref{app:proofs} for a proof.

\subsection{Estimating the FDOS using DQC1 circuits}
\label{sec:dos_with_circuits}

\subsubsection{One Clean Qubit (DQC1) computation}
\cref{eqn:FDOS_definition_trivial} may be rewritten
in terms of the maximally mixed state $\rho_{\text{max}}=\mathds{1}/d$ as
\begin{equation}
	G(t)=\frac{d}{\sqrt{2\pi}}\Tr[\rho_{\text{max}}e^{-i\hat{H}t}].
\end{equation}
Several related, prior works~\cite{ghanem2023robust,schuckert2023probing,lu2021algorithms}
estimate the FDOS based on the above equation, by sampling random pure initial states (given the average of all random initial basis
states is the maximally mixed state). For deriving our new results in the following sections, it is essential that here we consider another approach, whereby the FDOS can be computed using a purification of the maximally mixed state \cite{ferris2023exploiting}. When $\hat{H}$ is an $n$-qubit Hamiltonian, we add an $n$-qubit
ancillary register and create Bell pairs between the two registers as illustrated in \cref{fig:circ}(a)
-- indeed the reduced density matrix of the main register is the maximally mixed state, while the total system is in a pure state.
We then perform a Hadamard test, with the time evolution operator controlled on an ancilla qubit, 
and thereby deterministically encoding the FDOS into
the ancilla qubit's amplitude -- similar circuits have been investigated in the literature in the context
of the one clean qubit model (DQC1)~\cite{knill1998power}
 and thus we will refer to the approach as a DQC1 computation.
Finally, repeatedly measuring the ancilla qubit allows us to estimate the FDOS under standard shot noise scaling, however,
one could indeed apply
amplitude estimation to obtain a fundamentally improved precision -- although at the cost of a significantly increased circuit depth.

\subsubsection{Modified DQC1 computation for general subspaces \label{sec:dqc_subspace}}
\begin{figure}
    \centering
    \includegraphics[width=0.9\columnwidth]{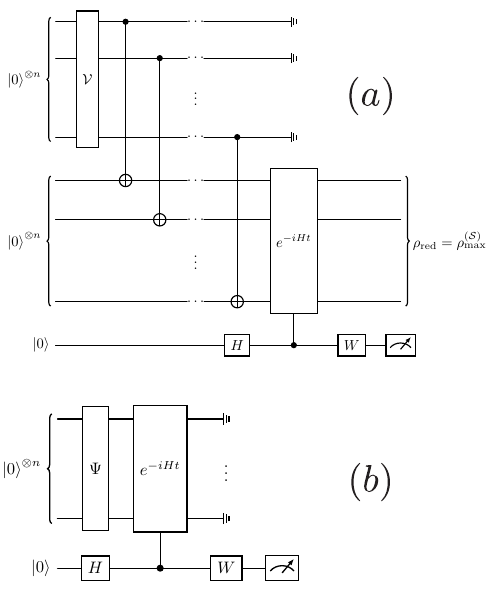}
    \caption{\textbf{Quantum circuits for evaluating the FDOS.}
    \textbf{(a)} DQC1 circuits:
      First, setting $\mathcal{V}$ to be the Hadamard transform $H^{\otimes n}$ allows us to
      estimate the FDOS over the entire Hilbert space  as relevant for \cref{sec:constant_particle_numb}.
      This way Bell pairs are created between the two registers
      such that the reduced density matrix of the lower, main register is maximally mixed -- as in standard DQC1 computations.
      Second, setting $\mathcal{V}$ to be a circuit that prepares a uniform superposition $|\mathcal{S}|^{-1/2} \sum_k \ket{\psi_k}$
      allows us to estimate the FDOS $G_\mathcal{S}(t)$ on a subspace (\cref{sec:variable_particle_numb}) spanned by $\{\ket{\psi_k}\}_{k\in \mathcal{S}}$.
      We detail an efficient construction of $\mathcal{V}$ for fermionic Hamiltonians in \cref{sec:dqc_subspace}.
      \textbf{(b)} Alternatively, the FDOS can be estimated by random state sampling methods outlined in \cref{sec:dos_with_random_states}.
      Here the main register is initialised in a randomly chosen initial state and and thus the need for a second register to prepare a maximally mixed state is eliminated entirely. The random intialisation circuit $\Psi$ can be extremely simple - even random single-qubit bit-flips suffice, but we explore further variants in \cref{sec:state_sampling_numerics}.
      The real (imaginary) parts of the FDOS are encoded as the ancilla probability by setting $W{=}H$ ($W{=}S^\dagger H$).
    }
    \label{fig:circ}
\end{figure}

Let us now consider a subspace spanned
by a set of states $\{\ket{\psi_k}\}_{k\in \mathcal{S}}$ that have a fixed particle number as relevant in \cref{sec:variable_particle_numb}.
The FDOS is then obtained as
\begin{equation}
    G_\mathcal{S}(t)=\frac{|\mathcal{S}|}{\sqrt{2\pi}}\Tr[\rho^{(\mathcal{S})}_{\text{max}}e^{-i\hat{H}t}].
    \label{eqn:FDOS_definition_subspace}
\end{equation}
Here the (normalized) projection of the maximally mixed state onto the subspace
is then obtained as
\begin{equation}
    \rho^{(\mathcal{S})}_{\text{max}}\equiv \frac{1}{|\mathcal{S}|}\sum_{k\in \mathcal{S}}\ket{\psi_k}\bra{\psi_k}.
\end{equation}
\begin{result}
The subspace variant $G_\mathcal{S}(t)$ can be computed via a simple modification
to the DQC1 protocol depicted in \cref{fig:circ}(a),
whereby we replace the Hadamard transform with a sub-circuit $\mathcal{V}$
that produces a uniform superposition of
all orthogonal states $\{\ket{\psi_k}\}_{k\in \mathcal{S}}$ in the subspace.	
\end{result}

The complexity of constructing the circuit $\mathcal{V}$ may strongly depend on the simulated
physical systems, however, for the practically pivotal case of fermionic problems under the Jordan-Wigner encoding
we can explicitly construct these circuits. 
For these systems, computational basis states have well-defined particle number equal to their Hamming weight;
the uniform superposition of fixed-Hamming-weight states is known as a Dicke state,
and indeed $\mathcal{V}$ is  a Dicke-state preparation circuit which can be implemented in
$\mathcal{O}(n)$ depth \cite{bartschi2019deterministic}. 
Indeed, as we show in the following section, the protocol can be performed by randomly preparing individual initial states spanning the subspace, rather than their uniform mixture --- a substantially easier task. This approach can therefore be used to estimate the DOS on any subspace for which the basis states can be efficiently prepared. This may be useful when considering symmetries and conserved quantities other than particle number --- for example, magnetic properties of a system might be elucidated by comparing the DOS on subspaces of fixed total spin $m_s$.

\subsection{Estimating the FDOS via random-state initialization}
\label{sec:dos_with_random_states}
Computing the FDOS via the circuit in \cref{fig:circ}(a) requires $2n{+}1$ qubits for an $n$-qubit Hamiltonian and
potentially a state-preparation circuit of depth $\mathcal{O}(n)$.
Similarly, previous works considered Monte Carlo importance sampling methods \cite{lu2021algorithms,schuckert2023probing,ghanem2023robust}
or approximating higher-order $3$-design random circuits~\cite{summer2024calculating} -- where the latter similarly introduces additional
circuit depth. In contrast, we now develop an approach that only requires $1$ ancilla qubit and introduces negligible overhead in circuit depth
through random-state sampling via a very straightforward $1$-design construction.

\begin{statement}\label{stat:unbiased_est}
We randomly choose from a unitary 1-design $\{U_i \}_i$
according to the probability distribution $p_i$ and thus we obtain a spherical 1-design 
as the states $|\phi_i\rangle = U_i |0\rangle$.
We define a random variable in terms of these random states as the Loschmidt echo at fixed time $t$ as
\begin{equation}
	\hat{L}(t) = \bra{\phi_i}e^{-i\hat{H}t}\ket{\phi_i} = \langle \phi_i(0) | \phi_i(t) \rangle,
\end{equation}
and obtain an unbiased estimator of the FDOS via 
\begin{align}
	\hat{G}(t) = d (2\pi)^{-1/2} \, \,     \hat{L}(t) ,
\end{align}
such that $\mathds{E}\left[   \hat{G}(t) \right] = G(t)$.
\end{statement}
Please refer to \cref{app:proofs} for a proof.
Above $\hat{L}(t)$ is the overlap between initial random state $\ket{\phi_i}$
and its time-evolved counterpart $\ket{\phi_i(t)}=e^{-i\hat{H}t}\ket{\phi_i}$
and can be computed via the Hadamard test circuit in \cref{fig:circ}(b).

 This replaces the initialization to the maximally mixed state in \cref{fig:circ}(a) with
initialization to $\ket{\phi_i}$ via the random circuit $\Psi$, 
which reduces the required number of qubits from $2n{+}1$ to $n{+}1$.
Indeed the above statement implies that unitary $1$-design
circuits are sufficient and allows	 one to forgo the significantly more complex
random circuits that approximate higher moments of the Haar distribution
used in previous works~\cite{PhysRevA.93.062306,summer2024calculating}.

A straightforward 1-design construction consists of applying bitflips with
$50\%$ probability to the individual qubits that are initialised to the $\ket{0}$ state and thus $\{ U_i\} = \{\openone,X\}^{\otimes n}$  and $p_i = 1/2^n$.
We numerically explore further 1-designs in \cref{sec:state_sampling_numerics}, such as continuous SU$(2)^{\otimes n}$ rotations,
and confirm that the choice of 1-design does not affect the convergence or shot requirements.

For estimating the DOS on a subspace, such as for fermionic or bosonic Hamiltonians,
computing the FDOS requires one to sample states uniformly within the relevant subspace. 
While in \cref{stat:unbiased_est} we assumed a 1-design property to make it clear how the present approach
improves upon previous 2- and 3-design constructions~\cite{PhysRevA.93.062306,summer2024calculating}, 
we now show that even less structure than a 1-design is sufficient for our approach
-- and this observation forms the basis for our generalisation to subspaces.
\begin{statement}\label{stat:subspace_sampling}
\cref{stat:unbiased_est} holds for any set of states $|\phi_i\rangle$
and probabilities $p_i$ that satisfy $\sum_i p_i | \phi_i \rangle\langle \phi_i| =  \rho_{\text{max}}$.
This allows us to naturally generalise our result to subspaces without requiring 1-design properties. In particular,
given a set of states within the subspace
$\{\ket{\psi_k}\}_{k \in S}$ and a probability distribution $p_i$ that satisfy the property
\begin{equation}
	\sum_{k \in S} p_i | \phi_i \rangle\langle \phi_i| =  \rho^{(\mathcal{S})}_{\text{max}},
\end{equation} 
then estimating the Loschmidt echo 
of randomly chosen initial states
$\hat{L}_\mathcal{S}(t) = \bra{\phi_i}e^{-i\hat{H}t}\ket{\phi_i}$
yields the unbiased estimator  for the FDOS as
\begin{align}
	\hat{G}_\mathcal{S}(t) = |\mathcal{S}|(2\pi)^{-1/2} \, \,     \hat{L}_\mathcal{S}(t),
\end{align}	
such that $\mathds{E}\left[  \hat{G}_\mathcal{S}(t)  \right]  = L_\mathcal{S}(t)$.
\end{statement}
Please refer to \cref{app:proofs} for a proof.
Again recall that fermionic Hamiltonians under the Jordan-Wigner encoding,
computational basis states have well-defined particle number $M$ equal to their Hamming weight. Thus, the set of computational basis states $\{\ket{k}\}_{k \in S}$ of fixed Hamming
weight sampled uniformly $p_i = |\mathcal{S}|^{-1}$ is a sufficient construction for our purposes.
More generally, one could initialise in a state of fixed particle number and
randomly apply number-state conserving transformations, such as matchgates \cite{wan2022matchgate},
to obtain a desired set of random initial states.

\begin{figure*}
	\centering
	\includegraphics[width=\textwidth]{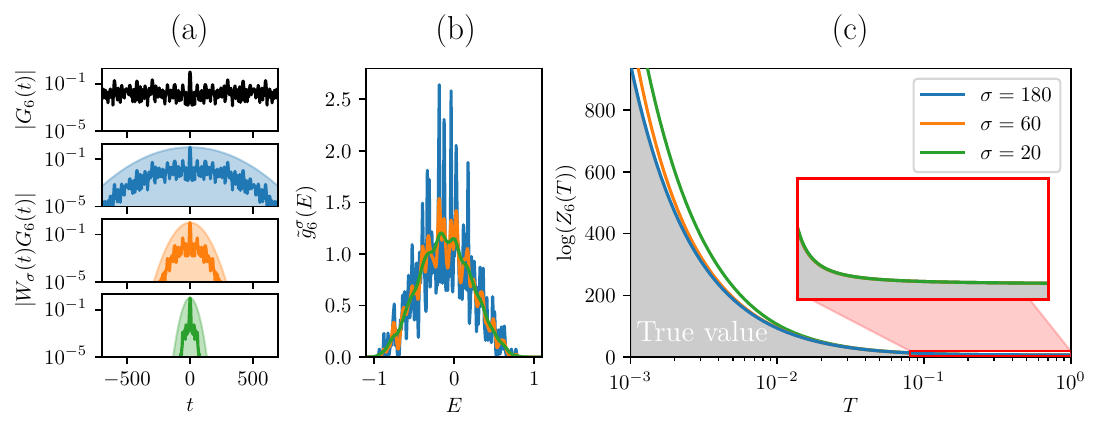}
	\caption{\textbf{Effects of finite simulation length.} 
	We study a $(3\times2)$ grid Fermi-Hubbard model 
	with open boundary conditions,
	nearest-neighbour couplings, $J=-1$ and $U=2$,
	and assuming the fixed-number subspace $M=6$.
	\textbf{(a)}
	The FDOS  $G_6(t)$ (black) is estimated using a quantum computer and
	allows us to reconstruct the DOS $g_6(E)$ on a subspace of fixed particle number $M=6$ through inverse Fourier transforming.
	As detailed in \cref{sec:window_width_effects}, to cap the total time evolution length, the signal is multiplied in post-processing
	by a Gaussian Fourier window $W_\sigma(t)$
	of temporal width $\sigma=180$ (blue), $\sigma=60$ (orange), and $\sigma=20$ (green)
	-- shaded regions represent the amplitude of the window function.
	\textbf{(b)} different resolution-limited DOS $\tilde{g}^{\sigma}(E)$
	are obtained from each signal in (a): wider temporal window width (blue) leads to
	finer-resolution spectral features while
	shorter time evolutions (green)  blur out spectral features.
	\textbf{(c)} the partition function $Z_6(T)$ from \cref{eqn:canonical_partition_function} is estimated for
	increasing temperatures $T=1/\beta$ using the resolution-limited DOS $\tilde{g}^{\sigma}(E)$ from (b)
	-- the shaded grey region represents the true partition function.
	Low-temperature physics imposes stricter requirements on the duration of dynamics:
	shorter temporal window widths (green and orange) yield poorer approximations at low temperatures due to
	the concentration of the Boltzmann factor $e^{-\beta E}$ at low energies.
	In contrast, the inset illustrates that even short-time dynamics can be sufficient 
	for accurate prediction of higher temperature properties.}
	\label{fig:spectral_resolution}
\end{figure*}

Finally, let us prove that the above sampling approaches are efficient as 
they admit a standard shot-noise scaling.
\begin{statement}	\label{stat:shot_number}
	 The number of samples required to estimate  $G_\mathcal{S}(t)/|\mathcal{S}|$ in \cref{stat:subspace_sampling} 
  (for the full Hilbert space $\mathcal{S} \equiv \mathcal{H} $ and
  $|\mathcal{S}| =d$)
	 to precision $\epsilon$ is upper bounded as $N_s \leq (2\pi)^{-1}  \epsilon^{-2}$.
	 Similarly, repeatedly measuring the ancilla qubit in our DQC1 circuits
	 \cref{fig:circ}(a) achieves the same bound.
\end{statement}
A proof is given in \cref{app:proofs}.
Importantly, our random initial state sampling approach in \cref{fig:circ}(b) achieves the same sample complexity
as our DQC1 circuits in \cref{fig:circ}(a), and therefore can be substituted for no additional cost (other than potential challenges in varying the executed circuit each shot, which is addressed in \cref{sec:state_sampling_numerics}). Although the maximally-mixed-state formulation is a theoretical convenience that is useful for proofs, explicitly preparing a maximally-mixed state on near-term devices is a challenging prospect that has hindered similar protocols~\cite{ferris2023exploiting}.
In contrast, trivial single-qubit operations (i.e. bit-flips) are sufficient for our approach, rendering the impact of improper state preparation negligible (i.e. preparation of states of correct particle number is assured to within single-qubit gate fidelity). Nonetheless, there may be advantages to using the more complex DQC1 circuits in a fault-tolerant setting.
For example, one could apply amplitude estimation to our DQC1 circuits
to estimate the ancillary probability 
which would enable a fundamentally improved time complexity, such that the runtime is $O(  \epsilon^{-1} )$
but in return the circuit depth becomes similarly $O(  \epsilon^{-1} )$.

\subsection{Reconstructing the DOS}
In the previous sections we detailed how to estimate the FDOS at a fixed time $t$ using a quantum computer.
While formally we can obtain the DOS by inverse Fourier transforming the FDOS as in \cref{eq:inv_trafo},
in practice we use a quantum computer to estimate $G(t)$ at discrete time intervals, and up to some finite time.

\subsubsection{Windowed time evolution}
\label{sec:window_width_effects}

In the present work we explore a windowed discrete inverse Fourier transform whereby we 
first estimate $G(t)$ in discrete steps of $\Delta t$ and multiply 
the resulting signal with a window function $W_{\sigma}(t)$ 
that has a characteristic temporal width $\sigma$. 
The inverse convolution theorem then guarantees
\begin{equation}
	\mathcal{F}^{-1}[W_\sigma(t)G(t)]=w_{1/\sigma}(E)\ast g(E),
	\label{eqn:blurred_density_of_states}
\end{equation}
where $\ast$ denotes convolution $f(E)\ast h(E)=\int d\epsilon f(\epsilon) h(E-\epsilon)$ and $w_{1/\sigma}(E)=\mathcal{F}^{-1}[W_\sigma(t)](E)$ is the inverse Fourier transform of the window function. We note that for typical window functions, $w_{1/\sigma}(E)$ is a point-spread function with characteristic energy width $1/\sigma$, and therefore \cref{eqn:blurred_density_of_states} represents the evaluation of the DOS up to a broadening of lines (i.e. `blurring' of the DOS). We will refer to
\begin{equation}
	\tilde{g}^{\sigma}(E)\equiv w_{1/\sigma}(E)\ast g(E)
\end{equation}
as the \emph{resolution-limited DOS}. 
While a broad range of window functions are available in the literature
that make different tradeoffs~\cite{harris1978use}, in
this work we consider a standard Gaussian window of width $\sigma$
and unit normalisation $\int dE w_{1/\sigma}(E)=1$ as
\begin{equation}
	W_\sigma(t)=\frac{1}{\sqrt{2\pi}}\exp(-\frac{t^2}{2\sigma^2}).
	\label{eqn:gaussian_window}
\end{equation}
In \cref{fig:spectral_resolution}(a) we plot our subspace FDOS signals for 
a Fermi-Hubbard Hamiltonian for increasing Gaussian window widths
and illustrate in \cref{fig:spectral_resolution}(b) how the 
spectral features are blurred out in the 
 resulting resolution-limited DOS $\tilde{g}_6^{\sigma}(E)$ reconstructions.

In complete generality, the resolution of the DOS improves with an increasing window width $\sigma$. However, it requires longer-duration time evolution to evaluate $G(t)$ in the sense that
resolving spectral features in the DOS to an energy scale $\Delta E=\mathcal{O}(1/\sigma)$
necessitates evolution up to a timescale $\mathcal{O}(\sigma)$.
This is indeed the main limitation of the present approach: the complexity of simulating
the time evolution using a quantum computer increases with the total simulation time
and can be rather challenging even for relatively short timescales using NISQ and early fault-tolerant devices.
Still, relatively shallow circuits can be very useful in estimating certain thermodynamic properties, e.g.,
determining qualitative features may  be possible using short temporal window widths,
such as inspecting for the presence of an insulator gap.

We note that unlike kernel polynomial methods \cite{weisse2006kernel,summer2024calculating}, there is no requirement to rescale the Hamiltonian such that its spectrum lies in a certain range, which would require some prior knowledge of the spectrum to reliably perform.
Instead, the range and resolution of energies are determined by the duration of dynamics
$t_{\text{max}}$ and the sampling period $\Delta t$.
In particular, sampling $t$ in $G(t)$ via a uniform grid of $N_t=t_{\text{max}}/\Delta t$ points,
yields a uniform grid of energies between $-\frac{\pi}{N_t}+\frac{2\pi}{N_t \Delta t}$ and $\frac{\pi}{\Delta t}$;
The range of energies can then be increased by decreasing $\Delta t$
while the spectral resolution can be improved by increasing the duration of evolution.

Although we employ a uniform spacing of simulation times throughout this work, a non-uniform spacing can also be used in principle --- this may be useful e.g. if one has obtained some prior knowledge of the range in which the eigenenergies are concentrated from some approximate classical means. This could enable the use of well-established advanced techniques in signal processing, such as compressed sensing. One particularly appropriate generalization of this type is the use of Monte Carlo importance sampling methods to estimate the windowed Fourier transform in \cref{eqn:blurred_density_of_states}, i.e. one samples the FDOS $G(\hat{t})$ at random times
$\hat{t}$ which are drawn from a probability distribution
$P(t)=W_\sigma(t)/\{ \int_{-\infty}^\infty dt W_\sigma(t)\}$.
Such an approach leads to the same spectral resolution limits and to the same shot-noise
properties given the FDOS is anyway estimated through random sampling -- 
an advantage of the Monte-Carlo approach is that it allows evaluation of $\tilde{g}^{\sigma}(E)$
for a continuous $E$ in principle. Importance sampling from a window function is commonly applied in
other Fourier-limited methods for computing spectral properties \cite{wang2023quantum}.

\subsection{Directly estimating thermodynamic properties}
We can use our resolution-limited DOS from the previous section to directly estimate
thermodynamic properties via \cref{eqn:canonical_partition_function}.
In \cref{fig:spectral_resolution}(c) we illustrate that
a relatively short temporal window width $\sigma$ does not significantly hinder
estimating  thermodynamic properties at relatively high temperatures (small $\beta$ via $T=1/\beta$)
as it primarily leads to inaccuracies at low temperatures  (large $\beta$).
This is of course expected from \cref{eqn:canonical_partition_function} as for lower temperatures (larger $\beta$)
the Boltzmann factor $e^{-\beta E}$ becomes more evenly spread out in $E$ leading to a diminished relevance
of small fluctuations in the density of states when calculating $Z(\beta)$. 

However, as we now show, thermodynamic properties in the canonical ensemble 
define effective window functions, which may be combined with importance sampling
to directly compute these properties -- eliminating approximations due to convolution
in the resolution-limited DOS.
Formally, thermodynamic properties are estimated via the integral
\begin{equation}
    F(\beta)=\int  f_\beta(E)g(E) \, \mathrm{d}E,
\end{equation}
where $f_\beta(E)$ is a function that encodes the desired property. 
For example, (a) one may directly compute the partition function $F(\beta)=Z(\beta)$
by setting $f_\beta(E)=e^{-\beta E}$ or
(b) the internal energy $F(\beta)=U(\beta)$ is obtained by setting $f_\beta(E)=Ee^{-\beta E}$.

Parseval's theorem allows us to apply
the Fourier transform to both functions without changing the result of the integral as
\begin{equation}
    F(\beta)=\int  \mathcal{F}[f_\beta(E)](t)G(t) \, \mathrm{d}t.
    \label{eqn:parseval_direct_computation}
\end{equation}
This expression can be estimated using importance sampling:
one randomly chooses times $\hat{t}$ according to the
probability density function as $P(\hat{t})=\mathcal{F}[f_\beta(E)](t)/\{\int_{-\infty}^\infty dt \mathcal{F}[f_\beta(E)](t)\}$,
which can be straightforwardly computed numerically and for relatively
simple functions $f_\beta(E)$ analytically. 
We note that, of course, a potential difficulty may be that the Fourier transform $\mathcal{F}[f_\beta(E)](t)$ 
has a long tail, e.g., the function may decrease asymptotically as $O(1/t)$ for increasing $t$, 
 in which case large evolution times $\hat{t}$ may be required beyond capabilities of near-term hardware.

\section{Numerical experiments}
\label{sec:numerics}
Here we demonstrate and benchmark our approach to computing the DOS in a variety of numerical experiments.
By performing explicit simulations of our protocols, including full-density-matrix simulations of realistic noise models, shot noise, and specific circuit decompositions, we demonstrate the practicality of our approach. In particular, we find that our methods require only very simple initial states (\cref{sec:state_sampling_numerics}), are extremely robust to both algorithmic error (\cref{sec:algorithmic_error}) and gate noise (\cref{sec:gate_noise}), and can even be made compatible with near-term quantum computers via a new technique for variational time evolution (\cref{sec:nisq_dos}).

\subsection{Random state sampling requirements}
\label{sec:state_sampling_numerics}
In \cref{sec:dos_with_random_states}, we concluded that our random initial state sampling approach in \cref{fig:circ}(b)
allows us to compute the FDOS using a comparable number of shots (circuit repetitions) as
the more complex DQC1 circuits in \cref{fig:circ}(a). Here we numerically compare our procedure to
when using four different sampling methods for estimating the FDOS as
\begin{itemize}
    \item measuring the ancilla qubit in DQC1 circuits
    \item exact Haar-random initial state sampling (which would be exponentially expensive)
    \item unentangled state sampling by applying single-qubit random rotations
    $e^{-i\theta_{m,1}\sigma^x_m}e^{-i\theta_{m,2}\sigma^z_m}e^{-i\theta_{m,3}\sigma^x_m}$ to each qubit $m$ with randomly chosen $\theta_{m,s}$ (Euler angles)
    \item computational basis state sampling by uniform randomly applying bitflips $\{\openone,X\}$ to each qubit on the computational zero state
\end{itemize}
Our focus in this section is to confirm convergence of the different approaches and verify that they indeed yield
the same result (up to possibly slightly different levels of shot noise)
-- given full Haar-random sampling is undefined for our subspace methods, we choose a model system 
as the Heisenberg chain
\begin{equation}
 \hat{H}=-J\sum_{j=1}^{n-1}\vec{\sigma}_j\cdot\vec{\sigma}_{j+1}+\sum_j^n h_j\sigma_j^z,
 \label{eqn:heisenberg_hamiltonian}
\end{equation}
and estimate the DOS in the full Hilbert space.
Here $\vec{\sigma}_j=[\sigma_j^x,\sigma_j^y,\sigma_j^z]$ is a vector of Pauli $x$, $y$ and $z$ matrices on the $j$th qubit, $J$ is a coupling constant,
and $h_j\in[-h,h]$ is sampled from a uniform distribution (for disorder strength $h$). As a metric for comparison, we define the error in the approximated 
DOS as
\begin{equation}
    \epsilon(f,\sigma)\equiv 1-\frac{\langle f,\tilde{g}^{\sigma}\rangle}{\sqrt{\langle f,f \rangle \langle \tilde{g}^{\sigma}, \tilde{g}^{\sigma} \rangle}},
\end{equation}
where $\langle f,g \rangle = \int dE f(E)g(E)$ is the $L^2$ inner product, and $f(E)$ is the density of states we estimate using $N_s$ shots.

\begin{figure}
    \centering
    \includegraphics[width=0.48\textwidth]{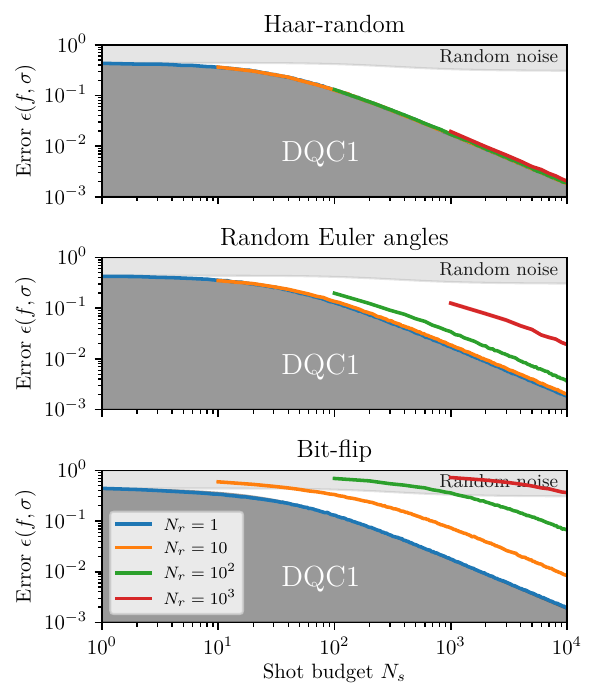}
    \caption{\textbf{Comparison of random initial-state sampling methods.} Error $\epsilon(f,\sigma)$
    	as a function of per-timestep shot budget $N_s$ in the DOS of a Heisenberg model in \cref{eqn:heisenberg_hamiltonian}
    	using different random sampling methods (Haar-random, single-qubit continuous rotations, and random bit flips).
    	As  in most hardware platforms loading a new circuit is expensive and thus multiple repetitions of a single circuit is desirable, 
    	 (coloured lines) represent that the same random initial state was reused $N_r$ times.
    	Lower dark grey shaded regions represent the errors obtained by DQC1 computation via \cref{fig:circ}(a)
    	whereas the light grey shaded area represents completely random noise.
    	Parallel lines in the above log-log plots confirm standard shot-noise
    		scaling  $N_s = O(\epsilon^{-2})$ from \cref{stat:shot_number} for all techniques, i.e.,
    		performances are only different by constant absolute factors, and for $N_r=1$ the random-sampling approach has the same sample complexity as the DQC1 computation as expected.
    		For larger values of $N_r$ (repeated initial
    		samples), the more uniform random sampling methods perform slightly better,
    		however,
    		they may require more complex initialisation circuits  -- the single-qubit, random Euler angle approach
    		is likely to be the most practical one.
    	Confidence intervals are too small to visualize on this plot.
    	}
    \label{fig:sampling_overlaps}
\end{figure}

While we investigate the effect of algorithmic errors and gate noise \cref{sec:algorithmic_error}, \cref{sec:gate_noise} \& \cref{sec:nisq_dos},
our focus here is to compare the performance of different sampling methods using exact time evolution.
In \cref{fig:sampling_overlaps} we plot the error $\epsilon(f,\sigma)$ for a range of total
per-timestep shot budgets $N_s$, and indeed confirm that all techniques
admit standard shot noise scaling (parallel lines in the log-log plot) as expected from our \cref{stat:shot_number}
-- although not visible on this
plot, we would expect the error $\epsilon(f,\sigma)$ to eventually plateau as we cap the maximal allowed time in $G(t)$
in our signal-processing approach as detailed in \cref{sec:window_width_effects}.

\cref{fig:sampling_overlaps} nicely illustrates that the DQC1 method requires slightly fewer shots
than random-state sampling (by a small absolute constant factor),
and that all three random-state sampling methods achieve the same performance when a new random
initial state is supplied for each shot (blue curves).
Indeed, there is no advantage in using complex random sampling methods, such as Haar-random sampling
as in Ref.~\cite{summer2024calculating},
as simple $1$-design samplings (e.g., bit-flip sampling) achieve the same performance
when a new initial state can be supplied for each shot.

However in practice, on most quantum hardware platforms it is expensive to load a different
circuit at each repetition, i.e. it is more efficient to `bunch' shots together by repeatedly sampling the same
initial state, and thus changing the circuit every $N_r$ shots. In this case, the total shot budget is
$N_s=N_r\times N_\psi$, where we sample $N_\psi$ random initial states.
\cref{fig:sampling_overlaps} (orange, blue, red) illustrates the performance when 
an initial random state is resued $N_r>1$ times. Surprisingly, Haar-random sampling performance
is almost uncompromised for up to $N_r =10^3$
while random Euler-angle single-qubit sampling admits a moderately increased shot noise
as we increase $N_r$.
In contrast, shot noise in bit-flip sampling is significantly increased for an increasing $N_r$.
This nicely illustrates that random-state sampling methods that sample the Hilbert space more uniformly can 
be advantageous as they may require fewer samples -- indeed, single-qubit Euler-angle sampling appears to be 
the most practical alternative.

Our motivation here for studying the convergence in a non-fermionic (fixed-particle-number) model
was the larger range of random-state sampling methods available which allowed us to showcase the effect
of uniformity in state sampling.
We repeat a similar experiment for the Fermi-Hubbard model in a strongly-interacting regime ($J=-1$, $U=8$) in \cref{appendix:sampling_convergence_hubbard},
drawing a similar conclusion and numerically demonstrating analogous convergence rates.

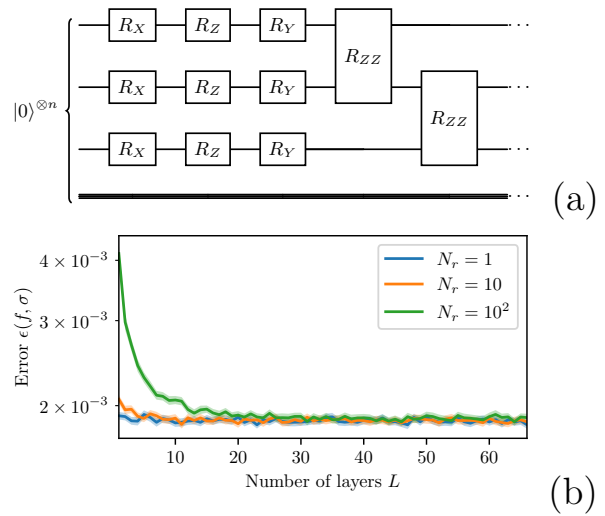
\begin{figure}
    \centering
    \begin{tikzpicture}
    \node[scale=0.8] {
    \begin{quantikz}
    \lstick[6]{$\ket{0}^{\otimes n}$} & \gate[1]{R_X} & \gate[1]{R_Z} & \gate[1]{R_Y} & \gate[2]{R_{ZZ}} & \qw & \dots\\
    & \gate[1]{R_X} & \gate[1]{R_Z} & \gate[1]{R_Y} & & \gate[2]{R_{ZZ}} & \dots \\
    & \gate[1]{R_X} & \gate[1]{R_Z} & \gate[1]{R_Y} & \qw & & \dots  \\
    \setwiretype{b} \qw & \qw & \qw & \qw & \qw & \qw & \dots
    \end{quantikz}
    };
    \end{tikzpicture} {\Large(a)}
    \includegraphics[width=0.4\textwidth]{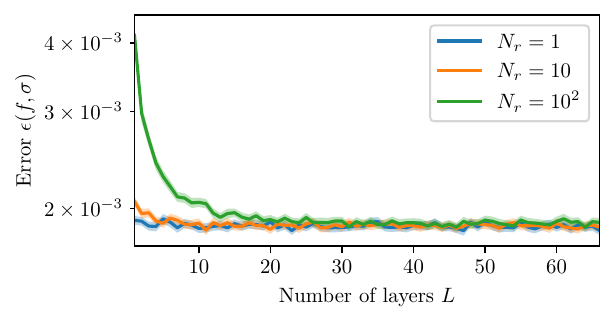} {\Large(b)}
    \caption{\textbf{Effect of initial-state-preparation circuit depth on DOS error.}
    	\textbf{(a)} We consider here random initial state sampling in \cref{fig:circ} through randomly choosing gate parameters
    	in the pictured layered ansatz. Each layer consists of arbitrary single-qubit rotations (parametrized by Tait-Bryan angles)
    	and nearest-neighbour $ZZ$ rotations.
    	\textbf{(b)} Error $\epsilon(f,\sigma)$ in the estimated DOS for an increasing
    	number of layers $L$, at three different sampling circuit repetitions $N_r\in\{1,10,10^2\}$.
    	Shaded bars represent bootstrapped 95\% confidence intervals.
    	Increasing $L$ interpolates between the $1$-design and $2$-design limits, the latter of which leads to reduced impact of
    	sampling circuit repetition. Indeed more uniform random sampling of the Hilbert space is preferable when
    	reusing initialisation circuits (via $N_r >1$).}
    \label{fig:layer_variation}
\end{figure}

To further explore this intuition that more uniform random-state sampling is advantageous when randomly-sampled initial states are repeated, we consider the use of a layered hardware-efficient ansatz circuit structure to generate random initial states by randomly
sampling ansatz parameters. The circuit, depicted in \cref{fig:layer_variation}(a), consists of $L$ layers each consisting of $X$-$Y$-$Z$ rotations (Tait-Bryan angles), followed by nearest-neighbour $ZZ$ couplings. It is well known that this circuit structure generates a full-rank Lie algebra, and thus generates an $\epsilon$-approximate $2$-design with layers $L\propto \log(1/\epsilon)$ \cite{larocca2021diagnosing}. Consequently, varying $L$ should vary between the $1$-design and $2$-design limits. In \cref{fig:layer_variation}(b), we compare the error $\epsilon(f,\sigma)$ achieved for $N_s=10^4$ shots per timestep at varying layer count $L$, for three different sample repetition numbers $N_r$. As expected, we see that in the limit of many layers $L$, the three schemes converge to nearly the same error $\epsilon(f,\sigma)$.

\begin{figure*}
	\centering
	\includegraphics[width=0.95\textwidth]{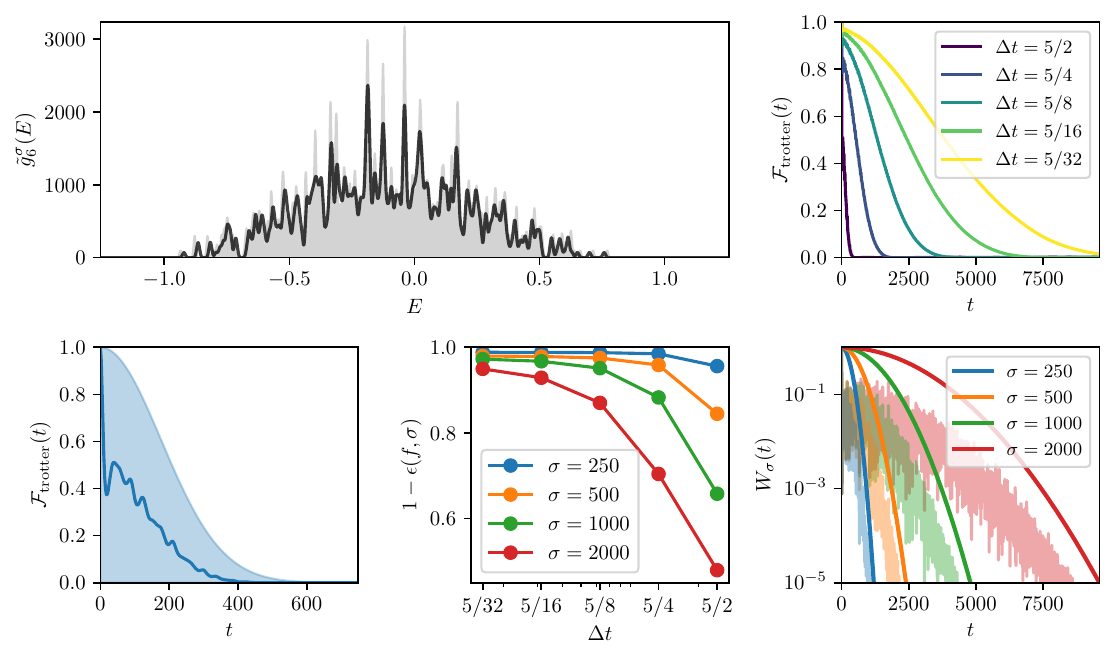}
	\caption{\textbf{Effect of algorithmic error on DOS calculations.} \textbf{Top left:} The DOS estimated with our approach (solid line) of the $M=6$ subspace
	of a $3\times 2$ grid Hubbard model ($J=-1$, $U=2$) is a surprisingly accurate reproduction of the the exact DOS (shaded grey region)
	despite very significant time evolution Trotter error, i.e.,
	the unitary fidelity $\mathcal{F}_{\text{trotter}}(t)$ (solid blue line, \textbf{bottom left panel})
	of the time evolution operator decays rapidly in the timescale set by the Fourier window (shaded region, bottom panel).
	Even with levels of Trotter error that would be too significant for most applications,
	the spectral content of the FDOS $G(t)$ remains accurate enough to correctly resolve the DOS.
	\textbf{Top right:} For a more systematic study, we compare five different Trotter steps
	$\Delta t\in\{5/2,5/4,5/8,5/16,5/32\}$ and depict the rate at which the unitary fidelity
	$\mathcal{F}_{\text{trotter}}(t)$ decays for each timestep $\Delta t$.
	\textbf{Bottom middle:} Using these timesteps $\Delta t$, we compare the error $\epsilon(f,\sigma)$ in the estimated DOS across a range of window widths $\sigma$. As expected, larger window widths $\sigma$ require shorter timesteps $\Delta t$, since these incorporate the effect of the signal $G(t)$ at later times $t$ where the accumulated effect of algorithmic error is greater. This is necessary to resolve more detailed spectral features, since the relevant spectral resolution is $\Delta E \sim 1/\sigma$. \textbf{Bottom right:} With respect to the timescale of the corresponding Fourier windows (solid lines) and windowed signals (transparent lines), the unitary fidelity $\mathcal{F}_{\text{trotter}}(t)$ decays non-negligibly, further confirming that our method is successful even with very significant algorithmic Trotter error.}
	\label{fig:algorithmic_error}
\end{figure*}

\begin{figure*}
	\centering
	\includegraphics[width=0.95\textwidth]{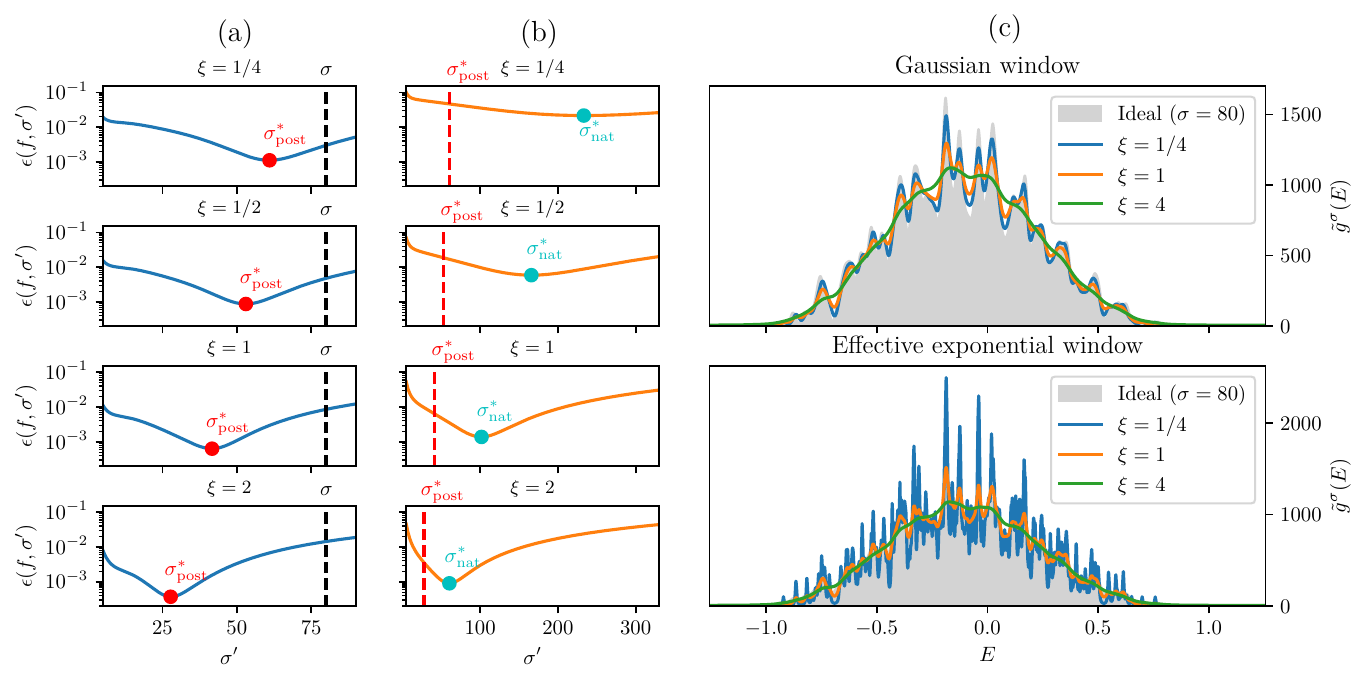}
	\caption{\textbf{Effect of hardware noise.} We compute the error $\epsilon(f,\sigma)$ in the DOS of a $3\times 2$ grid Hubbard model ($J=-1$, $U=2$) for $G_6(t)$ computed under the influence of an early fault-tolerant noise model at a range of different circuit
		error rates $\xi\in\{1/4,1/2,1,2,4\}$
		\textbf{(a)} We compare the estimated DOS (computed with fixed window width $\sigma=80$, black dotted line)
		to the ideal DOS that would be obtained with a perfect signal at window width $\sigma'$.
		We find that under the influence of noise, the error is minimized at a value of $\sigma'=\sigma^*_{\text{post}}$
		\emph{smaller} than $\sigma$, which decreases with increasing error rate $\xi$.
		This is due to the broadening-of-lines effect of gate noise.
		\textbf{(b)} Gate imperfections result in a natural effective exponential windowing.
		We compute the DOS using no additional windowing in postprocessing, and compare to the ideal DOS for an
		exponential window \eqref{eqn:exp_fourier_window} for a range of widths $\sigma'$.
		At each noise level, the (noise-induced, effective) window width $\sigma^*_{\text{nat}}$ that minimizes the error is larger than the corresponding optimal window width $\sigma^*_{\text{post}}$ obtained in (a) for the same dynamics, demonstrating the `natural' windowing induced by depolarizing noise in the dynamics.
		\textbf{(c)} Comparison of the DOS at differing error rates for Gaussian windowing (top) and no additional windowing (bottom). Increasing the error rate of the Trotter circuit leads to a broadening of lines in the DOS.}
	\label{fig:gate_noise}
\end{figure*}

\subsection{Resilience to algorithmic error}
\label{sec:algorithmic_error}
While simulating time evolution is possibly the most natural application of quantum computers, these
simulations are only approximate due to algorithmic errors. In this section, we implement 
Trotterized time evolution and investigate the effect of these algorithmic errors on the
DOS in the subspace of particle number $M=6$ as $g_6(E)$ -- we find surprising robustness to error even with low-fidelity evolution.

We consider the Fermi-Hubbard model from \cref{eqn:fermi_hubbard_hamiltonian}
for a $(3\times 2)$ grid with open boundary conditions,
nearest-neighbour coupling, and parameters $J=-1$, $U=8$, a strongly-interacting regime.
We map the Hamiltonian to qubits via a Jordan-Wigner encoding
and rescale the spectrum as in \cref{sec:window_width_effects}
to finally obtain $\hat{H}=\sum_m H_m$.
The time evolution under this Hamiltonian can be implemented via a 
first-order Trotterisation as
\begin{align}
	\nonumber
    U(\Delta t)&=\exp(-i\Delta t\sum_m H_m)\\
    &=    \prod_m \exp(-iH_m \Delta t)+\mathcal{O}(\Delta t^2),
    \label{eqn:first_order_trotter}
\end{align}
whereby we neglect all $\mathcal{O}(\Delta t^2)$ contributions.
Controlled time evolution is then implemented by applying controlled variants of the
individual local gates as $\ket{0}_a\bra{0}_a\otimes \openone + \ket{1}_a\bra{1}_a \otimes \exp(-iH_m \Delta t)$,
where $\ket{0}_a$, $\ket{1}_a$ denote ancilla states -- the resulting controlled Pauli rotations can be straightforwardly
implemented as we detail in \cref{appendix:ibm_eagle_model}.

We first compare the fidelity of the ideal dynamics to the Trotterized dynamics via
the usual unitary fidelity as
\begin{equation*}
	\mathcal{F}_{\text{trotter}}(t)\equiv\frac{1}{d}\Tr[\exp(i\hat{H}t)\prod_{s=1}^{t/\Delta t} \prod_m\exp(-iH_m\Delta t)].
\end{equation*}
\cref{fig:algorithmic_error}(top right) features the exponential drop in the fidelity as a function
of total simulation time $t$ for different choices of $\Delta t$.

We consider the modified DQC1 protocol outlined in \cref{sec:dos_with_circuits}, assuming the main register is in an initial maximally mixed state on the subspace of $M=6$ (which can be efficiently prepared via $\mathcal{O}(n)$-depth Dicke state circuits) and the ancilla is in the $\ket{0}_a$ state.
In \cref{fig:algorithmic_error}(top left), we show the evaluation of $g_6(E)$ for $\Delta t=2.5$ and a window width of $\sigma=250$.
Despite the very poor fidelity of trotter simulation---with $\mathcal{F}_{\text{trotter}}(t)$ in \cref{fig:algorithmic_error}(bottom left)
almost immediately dropping below $0.5$, and further decaying quickly relative to the the window
timescale $\sigma$---we find that the DOS is reproduced to a surprisingly good level of accuracy.
In fact, by comparing the individually resolved peaks in
\cref{fig:algorithmic_error} (top left, black vs grey lines) at the highest and lowest energies, it is clear
that the trotter error merely shifts the peaks to the left and right but does not significantly affect their
peak heights.

Indeed, according to the Baker--Campbell--Hausdorff formula, Trotterisation
implements the time evolution under the effective Hamiltonian 
$H_{eff} = \hat{H} + \Delta t P + O(\Delta t^2)$
where the perturbation operator $P$ consists of all commutators between the individual Hamiltonian terms.
As a consequence, the observed energies---as eigenvalues of the effective Hamiltonian $H_{eff}$---can only
deviate from the exact energies 
by at most as much as $\Delta E_k \leq \Delta t \lVert P \rVert_\infty + O(\Delta t^2)$.
These shifts can indeed be observed in \cref{fig:algorithmic_error} (top left)
as the window function used for convolving our DOS has a fine width $1/\sigma \ll  \Delta E_k$
narrower than the energy shifts.
However, in practice one should rather choose a convolution window (and hence total time evolution length)
such that $1/\sigma > \Delta E_k$ whereby the energy shifts due to Trotter error are not resolved -- 
this way the Trotter error effectively limits the achievable resolution.

In \cref{fig:algorithmic_error}(bottom middle), we compare the error $\epsilon(f,\sigma)$ achieved in the approximate DOS for varying window widths $\sigma\in\{250,500,1000,2000\}$, across a range of time-steps $\Delta t \in \{5/2, 5/4, 5/8, 5/16, 5/32\}$. Here we see another aspect to the relationship between window width and spectral resolution - as noted in \cref{sec:window_width_effects}, resolving spectral features in the DOS to an energy scale $\Delta E=\mathcal{O}(1/\sigma)$ necessitates evolution up to a timescale $\mathcal{O}(\sigma)$. However, accumulated algorithmic errors will increase in severity at larger $t$, necessitating smaller time-steps for accurate dynamics at later $t$. 
Nonetheless, we still find that the algorithmic error requirements are quite forgiving. Comparing the decay in unitary fidelities for these time-steps (\cref{fig:algorithmic_error}(top right)) to the timescales of the chosen window widths (\cref{fig:algorithmic_error}(bottom right)), we see again that reasonably accurate determination of the DOS is possible even with fidelities that decay fast within the window timescale $\sigma$, demonstrating the robustness of our scheme to Trotter error in particular.

\subsection{Resilience to gate noise}
\label{sec:gate_noise}
While algorithmic errors in the previous section can be suppressed through
increasing circuit depths, hardware and gate noise poses significant limitations on
the achievable circuit depths.
In this section, we consider the effect of gate noise on the Trotterised
methods studied in \cref{sec:algorithmic_error}, for an early fault-tolerant device.
As we detail in \cref{appendix:early_fault_tolerant_noise}, 
we assume that in an early fault-tolerant setting the dominant source of error will be
the imperfect implementation of continuous rotations (due to the relatively high cost of $T$ gates).
Our error model is parametrized by an overall circuit error rate $\xi=N_{\text{gates}}\lambda$
whereby $N_{\text{gates}}$ is the number imperfect, continuous-angle rotation gates
with fidelity $\lambda$.

As the logical qubits are effectively afflicted with Pauli noise \cite{kliuchnikov2023shorter,koczor2024sparse},
the state fidelity exponentially decays with increasing circuit depth to a very good approximation
as detailed in Ref.~\cite{foldager2023can}.
This acts as an effective exponentially-decaying Fourier window applied
to the temporal signals $G(t)$ as
\begin{equation}
    W_\sigma(t)=\frac{1}{\sqrt{2\pi}}\exp(-\frac{t}{\log(2)\sigma}),
    \label{eqn:exp_fourier_window}
\end{equation}
where the normalization has been chosen such that $\mathcal{F}^{-1}[W_\sigma(t)](E)=\frac{(1/\log(2)\sigma)}{\pi((1/\log(2)\sigma)^2+\omega^2)}$ is unit-normalized, ensuring correct normalization of the DOS. The effective window width $\sigma$ is determined by the strength of depolarizing noise and circuit structure.
As such, gate imperfections in trotter circuits yield a broadening of spectral features
and a reduction in spectral resolution $\Delta E$.
Indeed, this can be expected intuitively, the maximum number of gates 
is typically limited by the fact that $\xi=N_{\text{gates}}\lambda$ must not significantly
exceed one \cite{foldager2023can},
which imposes a limit on the duration of dynamics -- which indeed limits
the achievable spectral resolution.
Further reducing per-gate imperfections $\lambda$ in early fault-tolerant systems 
allow for an increased spectral resolution.

In \cref{fig:gate_noise}(a), we estimate the DOS at five different circuit
error rates $\xi\in\{0.25,0.5,1.0,2.0,4.0\}$,
and plot the deviation $\epsilon(f,\sigma)$ from the ideal windowed DOS 
using a range of different window widths $\sigma'$.
The unique minimum of the curves identifies the window width $\sigma^*_{\text{post}}$ that is most representative of the decay rate in the noisy DOS reconstructions --  and indeed $\sigma^*_{\text{post}}$ decreases as we increase $\xi$.

In \cref{fig:gate_noise}(b), we compute the DOS with no additional windowing, relying only
on the effective windowing introduced by gate imperfections,
and compare this to the ideal DOS to which an exponential window
from \cref{eqn:exp_fourier_window} is applied for increasing widths $\sigma'$.
Again, the unique minimum of the curves identifies the window width $\sigma^*_{\text{nat}}$
that is most representative of the noisy evolution, which in contrast to the previous example is naturally induced by gate noise. Similarly, $\sigma^*_{\text{nat}}$
decreases with $\xi$.
However, this window width is larger than that of the minimum in \cref{fig:gate_noise}(a),
indicating that spectral information is lost by additional windowing.
While some additional windowing will be required in the presence of shot noise,
the noise level should be taken into account when choosing a window width for postprocessing.

\subsection{Computing DOS on NISQ devices}
\label{sec:nisq_dos}
\begin{figure}
	\centering
	\includegraphics[width=0.5\textwidth]{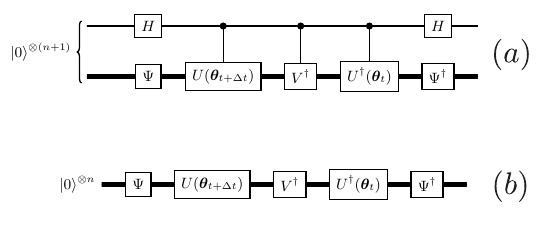}
	\caption{
		\textbf{Circuits for variational dynamics compilation.}
		The depicted circuits enable approximate recompilation of a Trotter step $V$ applied to some dynamical state $U(\bm{\theta_t})\Psi\ket{0}$, (a) such that $U(\bm{\theta}_{t+\Delta t})\Psi\ket{0}\approx VU(\bm{\theta}_t)\Psi\ket{0}$ including global phase or (b) such that $U(\bm{\theta}_{t+\Delta t})\Psi\ket{0}\approx e^{i\phi}VU(\bm{\theta}_t)\Psi\ket{0}$ for arbitrary global phase difference $\phi$. In both cases, the state produced by the depicted circuit minimizes the compilation Hamiltonian \eqref{eqn:compilation_hamiltonian} for the desired values of $\bm{\theta}_{t+\Delta t}$, enabling variational optimization at each time step.
		We use the CoVaR optimiser to train circuit parameters~\cite{boyd2022training}.
	}
	\label{fig:compilation_circuits}
\end{figure}

Error correction on current-generation devices is not yet of a sufficient scale to be useful for quantum dynamics.
In the NISQ era, we must contend with the underlying hardware errors of the physical gates themselves,
which are typically too severe even for simple dynamics methods like Trotterization
-- applying even as few as 20 Trotter steps for modest system sizes is at the edge
of current capabilities \cite{kim2023evidence}.
Substantial attention has been directed recently towards ansatz-based variational methods~\cite{cerezo2020variationalreview},
which have been applied to approximation of quantum dynamics
\cite{li2017efficient,cirstoiu2020variational,barison2021efficient,berthusen2022quantum,goh2023lie}.
While these methods vary in the details of their implementation, they typically use some ansatz
circuit $U(\bm{\theta})$ parametrized by a set of classical parameters $\bm{\theta}$, and attempt
to optimize the parameters at each time $t$ such that $U(\bm{\theta})\ket{\psi}\approx e^{-i\hat{H}t}\ket{\psi}$
on some initial state $\ket{\psi}$, possibly up to a global phase
(for our purposes the global phase is important since it becomes a relative phase under controlled evolution).
In \cref{appendix:variational_dynamics_covar}, we outline a new method of variational dynamics
using the circuits in \cref{fig:compilation_circuits} whose parameters are trained using the CoVaR approach~\cite{boyd2022training}.

Throughout this section, we evaluate all Loschmidt echoes under the effect of a realistic noise model based on the IBM Eagle processor, which is outlined in \cref{appendix:ibm_eagle_model}. This model is parametrized by a base noise multiplier $\lambda_0$, where $\lambda_0=1$ corresponds to noise levels comparable to current-generation hardware. We denote the combined channel containing both the circuit $U(\bm{\theta})$ and hardware noise as $\mathcal{U}_{\bm{\theta}}:\mathcal{D}(\mathcal{H})\to\mathcal{D}(\mathcal{H})$.

For this numerical experiment, we consider dynamics of the Heisenberg model in
\cref{eqn:heisenberg_hamiltonian} at $n=6$, for disorder strength $h=1$ and coupling $J=1$. This model has been identified as a strong candidate for early quantum advantage \cite{luitz2015many,childs2018toward}, and scaled up to larger system sizes would represent a genuinely non-trivial quantum task. We consider 100 initial states $\{\ket{\psi_s}\}_{s=1}^{100}$ sampled randomly from single-qubit rotations (uniform random Euler angles), and using the new variational method of \cref{appendix:variational_dynamics_covar}, compute approximate angles $\bm{\theta}_{t,s}$ such that $U(\bm{\theta}_{t,s})\ket{\psi_s}\approx e^{-i\hat{H}t}\ket{\psi_s}$ for each $s$ and relevant time $t$. We consider 60 timesteps of $\Delta t = 0.2$, evaluating each Loschmidt echo with shot noise of $N_s=200$ shots per time step (100 shots each for real and imaginary parts). We use a hardware-efficient ansatz outlined in \cref{appendix:variational_dynamics_covar}.
\begin{figure}
	\centering
	\includegraphics[width=0.48\textwidth]{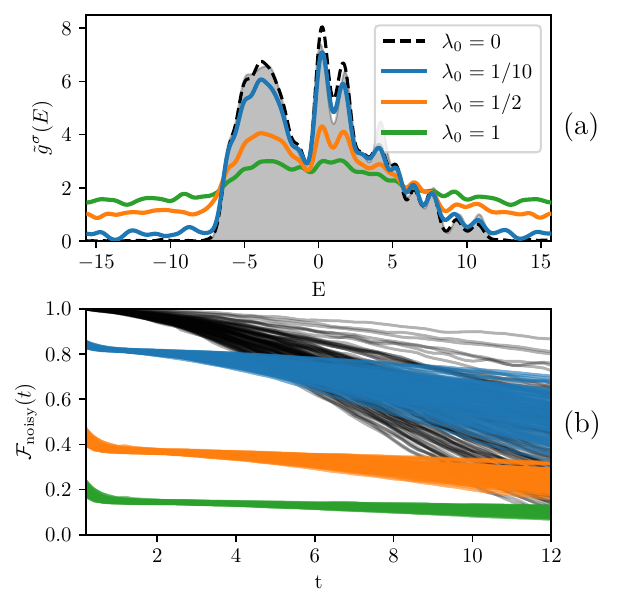}
	\caption{\textbf{Calculating density of states with noisy variational quantum dynamics.} \textbf{(a)} We demonstrate a NISQ-friendly variant of our methods by calculating the DOS of a Heisenberg model with disordered magnetic fields \eqref{eqn:heisenberg_hamiltonian} ($n=6$, $h=1$, $J=1$) in the presence of hardware noise, using variational quantum dynamics. We do not rescale the Hamiltonian, and use a window width of $\sigma=2.5$, leading to the ideal DOS depicted by the shaded grey region. 100 randomly sampled initial states are evolved with variational methods and used to estimate $G(t)$ at a per-timestep shot budget of $N_s=10^4$ ($N_r=100$), which in the absence of noise replicates the DOS with only minor defects (dotted black line). We compare the DOS obtained at varying overall noise levels $\lambda_0$, resolving the DOS to good fidelity at $\lambda_0=1/10$. \textbf{(b)} A comparison of the noisy fidelities $\mathcal{F}_{\text{noisy}}(t)$ for the underlying dynamics of the 100 randomly-sampled states, at each noise level.}
	\label{fig:variational}
\end{figure}

In \cref{fig:variational}, we compare this variational evaluation of the DOS for noise rates $\lambda_0\in\{1,1/2,1/10\}$. We compare the approximations to the DOS obtained in \cref{fig:variational}(a), finding that although noise comparable to current devices ($\lambda_0=1$) is too severe for the requirements of controlled evolution, by $\lambda_0=1/10$ we can reproduce a reasonable approximation of the DOS. This suggests that our methods may become more viable on NISQ devices with modestly increasing 2-qubit gate fidelity. In \cref{fig:variational}(b), we plot the state-to-state fidelities
\begin{equation}
    \mathcal{F}_{\text{noisy}}(t)\equiv\frac{1}{d}\Tr[\rho_{\text{ideal}}(t)\rho_{\text{actual}}(t)],
\end{equation} where $\rho_{\text{ideal}}(t)\equiv e^{-i\hat{H}t}\ket{\psi_s}\bra{\psi_s}e^{i\hat{H}t}$ is the ideal evolved state and $\rho_{\text{actual}}(t)=\mathcal{U}_{\bm{\theta}_{t,s}}(\ket{\psi_s}\bra{\psi_s})$ is the actual state obtained from the noisy ansatz.

\section{Conclusions\label{sec:conclusion}}

In this work we develop a range of new techniques for estimating thermodynamic properties
using quantum computers. The central object we consider is the Fourier transform of the 
DOS (FDOS) and develop new and improved methods for estimating it. First, we develop new methods
for estimating the DOS not only for the full Hilbert space but rather for a subspace
which is of crucial importance when simulating fermionic systems using quantum computers.
Second, we introduce deterministic circuits that enable direct sampling of the FDOS.
Third, we detail a broad range of random circuits that enable direct sampling of the FDOS
but using very simple 1-design initialisation circuits (random single qubit rotations are sufficient)
and controlled time evolutions.
We then detail that our FDOS sampling techniques enable the direct estimation of either the DOS or other
thermodynamic properties (through random FDOS importance sampling).

We simulate our approach in a broad range of numerical experiments. First, we analysed that
the approach is surprisingly robust against algorithmic as well as against gate errors that
are expected in early fault-tolerant quantum computers. We then introduced a novel variational
dynamics approach that uses the CoVaR optimiser to train parameters of a variational circuit~\cite{boyd2022training}
-- which then enabled us to demonstrate how the present approach may potentially be implemented in NISQ devices given there is no limit on how shallow we set the time evolution (affecting only the spectral resolution to which the DOS is faithfully resolved), and our techniques can further be combined with techniques like TE-PAI \cite{kiumi2024te}, which exchange depth requirements for sampling overhead.

Nonetheless, the ultimate value of NISQ devices for studying many-body quantum systems is currently a matter of contention \cite{zimboras2025myths} Indeed, the main bottleneck of our approach is the ability to perform time evolution, and setting the evolution time too short may lead to classical simulability. It seems increasingly likely that practically simulating dynamics beyond capabilities of classical methods will require circuit depths that necessitate some form of error correction. We believe that our approach may therefore be a practical enabler for the exploitation of early fault-tolerant devices, which are expected to enable medium to deep time evolution for practically relevant Hamiltonians beyond classical simulability.

As such, we believe an exciting future research direction is to assess resource requirements for performing DOS estimation beyond capabilities of classical computing. This will require establishing the following two components. First, one needs to establish boundaries of classical simulability by estimating the system size and time evolution depth where classical approximations become prohibitive. Second, one can estimate the quantum resources required for estimating DOS in two extreme limits. First, using Trotterisation and standard sampling requires fewer logical qubits and shallower quantum circuits, which relaxes code distance requirements and will therefore be well suited for early-fault tolerant implementations. Second, using our technique with an explicit maximally mixed initial state preparation combined with amplitude estimation will quadratically improve the previous runtime using standard sampling, ideally suited for implementation on
scalable fault-tolerant quantum computers. For the latter, fault-tolerant model we highlight that, while we focused on near-term friendly time evolution, such as Trotterisation, our approach is indeed compatible
with advanced time evolution algorithms, such as qubitisation which will enable exponentially better scaling with respect to algorithmic errors.

A range of further extensions and generalisations of our approach are also possible.
First, we concluded that our random initial state sampling approach
in \cref{fig:circ}(b) has the
same sample complexity as the more complex DQC1 circuits in \cref{fig:circ}(a),
however, the latter may still be advantageous in certain applications --
we noted that DQC1 circuits can be fundamentally improved through amplitude
estimation.
Second, the DQC1 circuits are compatible with techniques of \cite{yang2024phase} and
would thus enable one to estimate the FDOS
using only real and imaginary time evolutions thereby forgoing---the relatively more expensive---time evolutions
that are controlled on an ancilla qubit. In particular, one could estimate the FDOS $G(t)$
for a fixed $t$ applying the real and imaginary time evolutions to the mixed state in \cref{fig:circ}(a).
This may allow for trading off controlled time evolutions for a significantly increased number of shots -- 
as one estimates the phase via derivatives (with respect to the imaginary time) of the logarithm of Loschmidt echoes.
Another interesting future direction could be to explore whether using extrapolation
applied to increasing window sizes
can improve the quality of the DOS
(given sufficiently low error rates and shot noise).

\section*{Acknowledgements}
The authors thank Gregory Boyd, Hans Hon Sang Chan, Zo\"{e} Holmes, Tyson Jones, Ivan Rungger, and Fr\'{e}d\'{e}ric Sauvage for helpful technical conversations. We further thank Gregory Boyd for comments on the manuscript. MLG acknowledges the Rhodes Trust for the support of a Rhodes Scholarship. MLG was also (partially) supported by the Laboratory Directed Research and Development (LDRD) program of Los Alamos National Laboratory (LANL) under project number 20230049DR; and the Engineering and Physical Sciences Research Council under EPSRC project EP/Y004655/1.
BK thanks the University of Oxford for a Glasstone Research Fellowship and Lady Margaret Hall, Oxford for a Research Fellowship.
The numerical modelling involved in this study made
use of the Quantum Exact Simulation Toolkit (QuEST) \cite{jones2019quest} via the QuESTlink\,\cite{jones2020questlink} frontend. We are grateful to those who have contributed to all of these valuable tools. We thank Kristan Temme, Youngseok Kim and Andrew Eddins for supplying the parameters of the noise model outlined in \cref{appendix:ibm_eagle_model}.
The authors acknowledge the use of the University of Oxford Advanced Research Computing (ARC) facility\,\cite{oxfordARC} in carrying out this work.
BK thanks UKRI for the Future Leaders Fellowship Theory to Enable Practical Quantum Advantage (MR/Y015843/1).
The authors also acknowledge funding from the
EPSRC projects Robust and Reliable Quantum Computing (RoaRQ, EP/W032635/1)
and Software Enabling Early Quantum Advantage (SEEQA, EP/Y004655/1).

\clearpage
\newpage
\widetext
\appendix

\section*{Appendices for ``Direct Estimation of the Density of States for Fermionic Systems''}

\section{Proofs \label{app:proofs}}
\subsection{Proof of \cref{state:fdos}}
\begin{proof}
	\begin{align}
		G(t) := \mathcal{F}[g(E)](t) &= \frac{1}{\sqrt{2\pi}}\int_{-\infty}^\infty dE e^{-iEt}g(E) \\ &=\frac{1}{\sqrt{2\pi}}\sum_{k=1}^{d}  e^{-i E_k t}\\
		&= \frac{1}{\sqrt{2\pi}}\Tr[e^{-i\hat{H}t}],
	\end{align}
	where we have used the unitary, angular-frequency convention for the Fourier transform and the DOS definition of \cref{eqn:trivial_DOS}. We note for all time signals in this work, $G(t=0)=d/\sqrt{2\pi}$ and $G(-t)=(G(t))^*$, so only forward time evolution is required. 
	It follows that the DOS can be reconstructed via an inverse Fourier transformation as
	\begin{equation}\label{eq:inv_trafo}
		g(E)=\mathcal{F}^{-1}[G(t)](E)=\frac{1}{\sqrt{2\pi}}\int_{-\infty}^\infty dt e^{iEt}G(t).
	\end{equation}
\end{proof}

\subsection{Proof of \cref{stat:unbiased_est}}
\begin{proof}
	Expanding the expected value we obtain
	\begin{align}
		G(t)/d&= \mathds{E}\left[  \frac{1}{\sqrt{2\pi}}  \bra{\phi_i}e^{-i\hat{H}t}\ket{\phi_i} \right]\\
		&= \frac{1}{\sqrt{2\pi}}\sum_i p_i \bra{\phi_i}e^{-i\hat{H}t}\ket{\phi_i}\\
		&= \frac{1}{\sqrt{2\pi}}\sum_i p_i \tr [  e^{-i\hat{H}t}   \ket{\phi_i}\bra{\phi_i}] \label{eqn:proof_step}\\
		&= \frac{1}{\sqrt{2\pi}} \tr [  e^{-i\hat{H}t}   \sum_i p_i  U_i \ket{0}\bra{0}U_i^\dagger].
	\end{align}
	By definition a 1-design recovers Haar-random averages in the first moment, and thus
	\begin{align}
		&\frac{1}{\sqrt{2\pi}}  \tr [  e^{-i\hat{H}t}  \sum_i p_i  U_i \ket{0}\bra{0}U_i^\dagger]\\
		=& \frac{1}{\sqrt{2\pi}}   \tr [  e^{-i\hat{H}t}   \int_\psi   U \ket{0}\bra{0} U^\dagger \mathrm{d}U ]   \\
		=& \frac{1}{\sqrt{2\pi}}\tr [  e^{-i\hat{H}t} \rho_{\text{max}}],
	\end{align}
	where we used that the integration over the Haar measure yields the maximally mixed state $\rho_{\text{max}}$.

\end{proof}

\subsection{Proof of \cref{stat:subspace_sampling}}
\begin{proof}
	Indeed, we can re-prove \cref{stat:unbiased_est} by starting the same way as
	\begin{align}
		G(t)/d&= \mathds{E}\left[  \frac{1}{\sqrt{2\pi}}  \bra{\phi_i}e^{-i\hat{H}t}\ket{\phi_i} \right]\\
		&= \frac{1}{\sqrt{2\pi}}\sum_i p_i \bra{\phi_i}e^{-i\hat{H}t}\ket{\phi_i}\\
		&= \frac{1}{\sqrt{2\pi}}\sum_i p_i \tr [  e^{-i\hat{H}t}   \ket{\phi_i}\bra{\phi_i}],
	\end{align}
	but at this point we move the summation inside the trace as
	\begin{align}
				G(t)/d&=  \frac{1}{\sqrt{2\pi}}\sum_i p_i \tr [  e^{-i\hat{H}t}   \ket{\phi_i}\bra{\phi_i}]\\
				&=  \frac{1}{\sqrt{2\pi}} \tr [  e^{-i\hat{H}t}  \sum_i p_i \ket{\phi_i}\bra{\phi_i}].
	\end{align}
	By applying $\sum_i p_i | \phi_i \rangle\langle \phi_i| = \rho_{\text{max}}$ we obtain the desired expression.
	Similarly, for subspace approach we assume the property $\sum_i p_i | \phi_i \rangle\langle \phi_i| =  \rho^{(\mathcal{S})}_{\text{max}}$ thus we obtain the desired
	 expression for $G_\mathcal{S}(t)$.
\end{proof}

\subsection{Proof of \cref{stat:shot_number}}
The variance of the estimator is
\begin{equation}
	\var[G(t)/d] =  (2\pi)^{-1} \, \var\left[   \hat{L}(t) \right] \leq (2\pi)^{-1} \, \,  \mathds{E}\left[   |\hat{L}(t)|^2 \right],
\end{equation}
and given that $|\hat{L}(t)|^2 \leq 1$ we have $\var[G(t)/d] \leq (2\pi)^{-1}$.
Thus, the number of samples required to estimate  $G(t)/d$ to precision $\epsilon$ is upper bounded through the
variance of the mean estimator as $N_s \leq (2\pi)^{-1}  \epsilon^{-2}$. The variance of $G_\mathcal{S}(t)/d$ follows immediately from the same argument.

\section{Sampling convergence in subspaces}
\label{appendix:sampling_convergence_hubbard}
\begin{figure}
    \centering
    \includegraphics[width=0.48\textwidth]{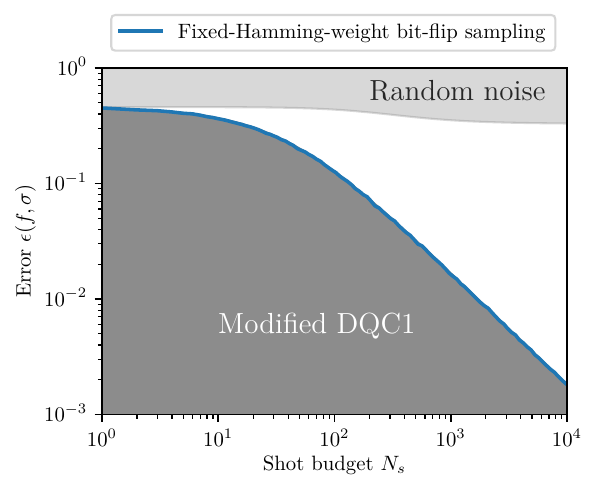}
    \caption{\textbf{Convergence of sampling methods for fermionic problems.} We plot the error $\epsilon(f,\sigma)$ as a function of per-timestep shot budget $N_s$ in the DOS of a Hubbard model \eqref{eqn:fermi_hubbard_hamiltonian} using the modified DQC1 approach for fermionic subspaces (black line and dark shaded region) and bit-flip initial state sampling at fixed Hamming weight (blue line), which again coincide. The upper light grey shaded area represents the error that would be obtained from a random white-noise signal. The behaviour is overall analogous to that of the non-fermionic problem in Figure \ref{fig:sampling_overlaps}. Confidence intervals are too small to visualize on this plot.}
    \label{fig:sampling_overlaps_hubbard}
\end{figure}

In Section \ref{sec:state_sampling_numerics}, we numerically studied the convergence of random-state-sampling methods for computing the FDOS, versus the DQC1 method, and found that both converged at a standard shot-noise-limited rate (i.e. parallel lines in the log-scale plot of Figure \ref{fig:sampling_overlaps}) when applied to a Heisenberg spin chain, for which `particle number' is not well-defined (i.e. it is not a fermionic problem). There, we focused on this problem due to the wider range of random-state sampling methods available, to highlight how the uniformity of sampling method affects the impact of circuit repetitions $N_r$. Here, we demonstrate that the same convergence properties also apply to fermionic problems, by performing the same sampling task on a Fermi-Hubbard model (\cref{eqn:fermi_hubbard_hamiltonian}). As in the rest of this work, we utilize a $(3\times 2)$ grid Hubbard model with open boundary conditions, nearest-neighbour coupling, and parameters $J=-1$, $U=2$. The DOS is computed with a window width $\sigma=60$ (relative to the rescaling described in Section \ref{sec:window_width_effects}). Sampling is performed (a) with the modified fermionic-subspace DQC1 approach of Section \ref{sec:dqc_subspace}, and (b) with fixed-Hamming-weight bit-flip sampling as outlined in Section \ref{sec:dos_with_random_states}. The results are depicted in \cref{fig:sampling_overlaps_hubbard}, again confirming standard shot-noise-limited scaling, but this time for a fermionic problem.

Beyond this simple bit-flip sampling approach, we anticipate that more uniform sampling could be obtained by e.g. initializing in a state of definite number and applying random (number-conserving) matchgate circuits \cite{wan2022matchgate}, yielding the benefits of more uniform sampling methods seen in Section \ref{sec:state_sampling_numerics} when $N_r>1$.

\section{Hardware error models}
\subsection{Early fault-tolerant devices}
\label{appendix:early_fault_tolerant_noise}
For our analysis of noise effects in early fault-tolerant devices in Section \cref{sec:gate_noise}, we apply a modified version of the relevant error model in Ref.~\cite{chan2022algorithmic}. Here, we assume that the dominant error source is the application of $T$ gates and that the error mechanism 
is due to the high cost of magic state distillation, which forces one to use
short-depth Clifford+T sequences in early-fault tolerance; This typically results in an effective depolarizing noise model~\cite{kliuchnikov2023shorter} --  which may be mitigated through more advanced techniques \cite{koczor2024sparse}. We compute time-evolution via first-order Trotterization \cref{eqn:first_order_trotter}, which for controlled evolution of the Hubbard model \cref{eqn:fermi_hubbard_hamiltonian} under a Jordan-Wigner mapping generates terms of the form
\begin{equation}
    \ket{0}_a\bra{0}_a \otimes \openone + \ket{1}_a\bra{1}\otimes \exp(-i\Delta t \sigma_i^\nu \sigma_{i+1}^z \dots \sigma_{j-1}^z \sigma_j^\nu),
    \label{eqn:jordan_wigner_rotations}
\end{equation}
with $\nu\in\{x,y,z\}$. The Coulomb terms of the Hubbard Hamiltonian act on a single site $i=j$, and the hopping terms generate more general terms of the form in \cref{eqn:jordan_wigner_rotations}. Although we do not directly model this, the $\sigma^z$ terms sandwiched between qubits $i$ and $j$ in the hopping terms can be removed by introducing a network of fermionic \textsc{swap} (\textsc{Fswap}) gates, which only consists of local gates of depth $\mathcal{O}(N^{\frac{1}{2}})$~\cite{PhysRevApplied.14.014059,PhysRevApplied.18.044064,PhysRevA.79.032316,PhysRevLett.120.110501}. This motivates our error model where local depolarizing noise of probability $\lambda$ is applied to the ancilla and qubits $i$ and $j$ for every controlled Pauli rotation. We parametrize this in terms of a total circuit error rate $\xi=\lambda N_{\text{gates}}$, where our Trotter circuit contains $N_{\text{gates}}$ error-susceptible gates due to imperfect application of $T$ gates.

\subsection{NISQ devices}
\label{appendix:ibm_eagle_model}
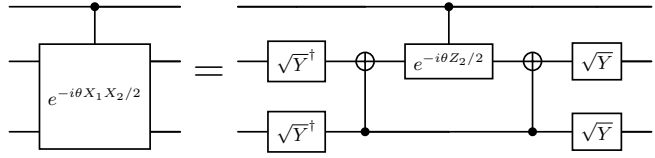
\begin{figure}
    \centering
    \begin{tikzpicture}
    \node[scale=0.8] {
    \begin{quantikz}
    \qw & \ctrl{1} & \qw \\
    \ghost{\sqrt{Y}^\dagger} & \gate[2]{e^{-i\theta X_1 X_2 / 2}} & \qw \\
    \ghost{\sqrt{Y}^\dagger} & & \qw
    \end{quantikz}
    {\huge =}
    \begin{quantikz}
        \qw & \qw & \qw & \ctrl{1} & \qw & \qw & \qw \\
        \qw & \gate{\sqrt{Y}^\dagger} & \targ{} & \gate{e^{-i\theta Z_2/2}} & \targ{} & \gate{\sqrt{Y}} & \qw \\
        \qw & \gate{\sqrt{Y}^\dagger} & \ctrl{-1} & \qw & \ctrl{-1} & \gate{\sqrt{Y}} & \qw
    \end{quantikz}
    };
    \end{tikzpicture}
    \caption{
        \textbf{Decomposition of controlled multi-qubit Pauli rotations.} Here we show a decomposition of a controlled-Pauli-XX rotation, whereby the entire operation can be controlled by controlling one single-qubit Pauli-Z rotation. The Pauli product is conjugated by single-qubit gates and CNOTs which map it to a single Pauli $Z$ observable, about which a phase rotation is applied. This can be straightforwardly generalized to any controlled multi-qubit Pauli rotation using conjugation by appropriate single-qubit transfomations and CNOTs, but always can be controlled via one single-qubit Pauli-Z rotation.
    }
    \label{fig:controlled_pauli}
\end{figure}

In \cref{sec:nisq_dos}, we explore DOS calculation in the NISQ era using variational methods. This necessitates a realistic noise model based on current-generation quantum hardware, which we detail here. Since our methods require controlled time-evolution, the overwhelming majority of gates for the circuits of \cref{sec:nisq_dos} are 2-qubit (from single-site Hamiltonian terms) or 3-qubit (from coupling Hamiltonian terms) gates. Here we outline a NISQ error model for these multi-qubit gates. First, we note that although simulating controlled evolution under the Heisenberg Hamiltonian \cref{eqn:heisenberg_hamiltonian} requires 3-qubit gates (controlled multi-qubit Pauli rotations), these can be easily be decomposed into 1- and 2-qubit gates by the structure outlined in \cref{fig:controlled_pauli}. Therefore, when applying controlled coupling terms, we apply a 2-qubit error channel between the two target qubits and a (different) two-qubit error channel between the ancilla and the nearest target qubit. Since qubit connectivity varies greatly between architectures, we do not explicitly model it here.

Each distinct pair of qubits has a unique error model, in order to capture the effects of differing calibration properties of each qubit. The error models, previously used in Ref.~\cite{ijaz2024more}, are drawn from a realistic noise model that was learned directly from a 127-qubit IBM Eagle processor in Ref. \cite{kim2023evidence}. The noise models are of the sparse Pauli-Lindblad form outlined in Ref. \cite{berg2022probabilistic}. Each channel is generated by a site-dependent Lindbladian $\mathcal{L}_m$, where
\begin{equation}
    \mathcal{E}_m(\rho)=\exp\left[\mathcal{L}_m\right](\rho),\quad \mathcal{L}_m(\cdot)=\lambda_0\sum_{k\in\mathcal{K}}\gamma_{km}(P_k\cdot P_k^\dagger-I\cdot I)
\end{equation}
for $\gamma_k\geq 0$, and where $\lambda_0$ is an overall multiplier of the base noise level and $\mathcal{K}$ is the set of 2-qubit Pauli operators for the qubits acted upon by the entangling gate. 

This corresponds to a diagonal Pauli transfer matrix with diagonal terms $f_j=\prod_{k\in\mathcal{K}}w_k+(1-w_k)(-1)^{\langle j,k\rangle}$, where $w_k\equiv(1+e^{-2\lambda\gamma_k})/2$ and $\langle a,b \rangle$ is the binary symplectic product
\begin{equation}
\langle a, b \rangle =  
\begin{cases}
0 \textrm{  when  } [P_a, P_b]=0,\\
1 \textrm{  when  } \{P_a, P_b\}=0,
\end{cases}
\label{eqn:sympl_product}
\end{equation}
One can easily write this in Kraus operator form
\begin{equation}
    \Lambda(\cdot)=\sum_j c_j P_j \cdot P_j^\dagger,
    \label{eqn:kraus_form}
\end{equation}
by the Walsh-Hadamard type transform
\begin{equation}
    c_b =\frac{1}{4^n}\sum_a (-1)^{\langle a,b \rangle}f_a.
\end{equation}
In our numerical simulations, we generate the Kraus-operator forms of our noise channels \eqref{eqn:kraus_form} for a given noise level $\lambda$ and apply these directly to the density matrix after each entangling gate. Different $\gamma_{km}$ are chosen for each qubit pairing $m$, sampled randomly from a set of amplitudes learned directly from CNOT processes at different sites in a 127-qubit IBM Eagle device \cite{kim2023evidence}.

\section{Variational dynamics}
\label{appendix:variational_dynamics_covar}
To variationally approximate the dynamics in \cref{sec:nisq_dos}, we choose a parametrized ansatz circuit $U(\bm{\theta})$. Given an initial state $\ket{\psi}$ (produced by some sampling circuit $\Psi$ such that $\ket{\psi}=\Psi\ket{0}$), we seek to find parameters $\bm{\theta}_t$ such that $U(\bm{\theta}_t))\ket{\psi}\approx e^{-i\hat{H}t}\ket{\psi}$. That is, $U(\bm{\theta}_t)$ should approximately reproduce the effects of time evolution \emph{on a specific initial state $\ket{\psi}$}, \emph{including the correct global phase}. This is a stronger requirement than typical variational dynamics schemes \cite{li2017efficient,cirstoiu2020variational,barison2021efficient,berthusen2022quantum,goh2023lie} due to the global phase requirement, but a much more permissive requirement than full unitary compilation $U(\bm{\theta}_t)\approx e^{-i\hat{H}t}$, which would require the effects of time evolution to be reproduced on \emph{all} initial states.

Here we introduce a novel form of variational approximation to quantum dynamics. Like many related methods \cite{barison2021efficient,berthusen2022quantum}, our approach is based on recompiling Trotter steps. At $t=0$, we initialize our parameters $\bm{\theta}_{t=0}$ such that $U(\bm{\theta})=\openone$ (typically this is $\bm{\theta}_{t=0}=\bm{0}$). Then, for each subsequent step, given parameters $\bm{\theta}_t$ at some time $t$, we attempt to find parameters for the next time $t+\Delta t$ such that
\begin{equation}
    U(\bm{\theta}_{t+\Delta t})\ket{\psi}\approx V(\Delta t)U(\bm{\theta}_t)\ket{\psi},
\end{equation}
where $V(\Delta t)\approx e^{-i\hat{H}\Delta t}$ is a Trotter step over time $\Delta t$, and global phase is accounted for in the (approximate) equality. This can be accomplished by applying the circuit depicted in Figure \cref{fig:compilation_circuits}(a), and variationally minimizing the compilation Hamiltonian
\begin{equation}
    H_{\text{comp}}=-\sum_{j=1}^{n+1}\sigma_j^z
    \label{eqn:compilation_hamiltonian}
\end{equation}
with respect to $\bm{\theta}_{t+\Delta t}$. The global phase is accounted for by the $n+1$th (ancilla) qubit. We note that in cases where global phase is unimportant, one could instead apply the circuit in \cref{fig:compilation_circuits}(b), which does not require controlled evolution on an ancilla, and thus $H_{\text{comp}}=-\sum_{j=1}^n \sigma_j^z$. Our approach can therefore be directly compared to related methods based on Trotter step recompilation. The optimization steps of such methods can be relatively costly, relying on either SWAP tests or gradient computation with parameter-shift rules \cite{barison2021efficient,lin2021real,berthusen2022quantum} which can have unfavourable sampling costs. We instead train our circuits with the CoVariance Root finding (CoVaR) approach introduced in Ref.~\cite{boyd2022training}. This allows very shot-frugal optimization of variational dynamics: not only do we not require a SWAP test or expensive gradient computation, but since only covariances of local observables are necessary, each step can be performed simply by measuring a classical shadow \cite{huang2020predicting}, necessitating a sample complexity that is merely logarithmic in $n$ for fixed observable locality. Furthermore, we note that since $\Delta t$ will typically be chosen to be small (such that the Trotter approximation holds), $\bm{\theta}_t$ and $\bm{\theta}_{t+\Delta t}$ will be close to each other in parameter space. This ensures that CoVaR is initialized in the a vicinity of the solution, where it performs best, and also may help circumvent the barren plateaus that plague variational methods in general \cite{barison2021efficient}.

\begin{figure}
    \centering
    \includegraphics[width=0.9\textwidth]{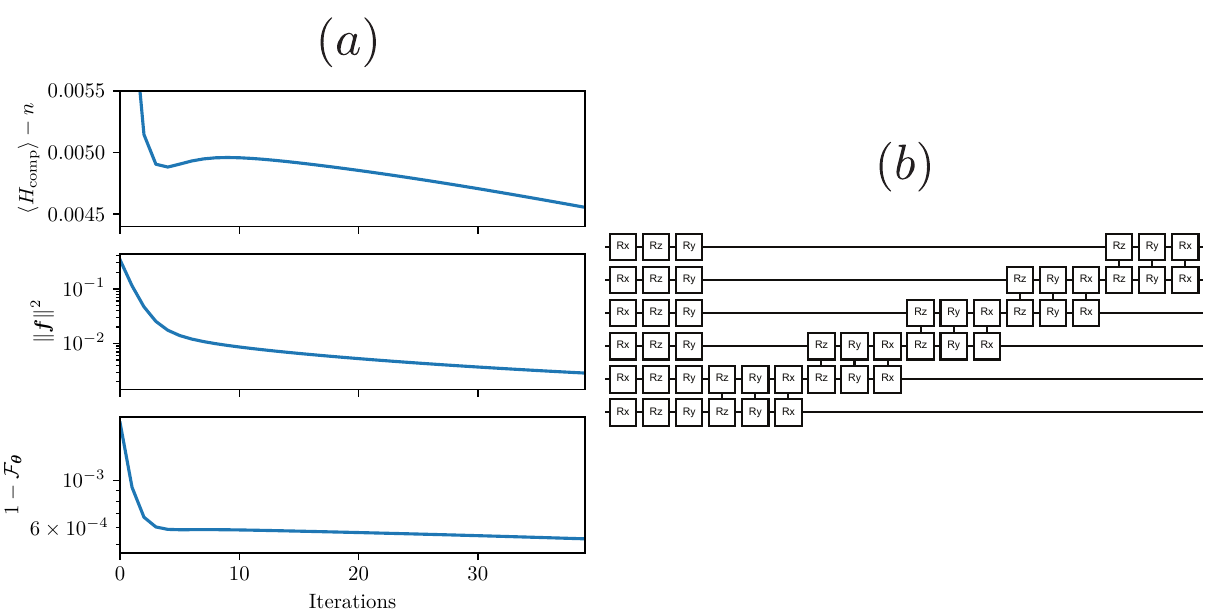}
    \caption{\textbf{Recompiling Trotter steps with CoVaR.}  \textbf{(a)} We recompile a single Trotter step of $\Delta t=0.2$ for the Heisenberg chain \eqref{eqn:heisenberg_hamiltonian}, with parameters as studied in \cref{sec:nisq_dos} ($n=6$ spins, disorder $h=1$, coupling $J=1$) and 40 CoVaR steps (one classical shadow required each). We compare the compilation energy $\langle H_{\text{comp}} \rangle$, the covariance norm $\|\bm{f}\|^2$, and the step infidelity $1-\mathcal{F}_{\bm{\theta}}$. The covariance norm $\|\bm{f}\|^2$ is directly minimized by the method, which leads here to a strict decrease in infidelity $1-\mathcal{F}_{\bm{\theta}}$. However, even though it strictly improves fidelity, this optimization takes a path through parameter space that briefly \emph{increases} energy $\langle H_{\text{comp}} \rangle$, which gradient-based methods would not do. \textbf{(b)} A single layer of the hardware-efficient ansatz used for variational approximation of dynamics in (a) and \cref{sec:nisq_dos}.}
    \label{fig:covar_trap_escape}
\end{figure}

In \cref{fig:covar_trap_escape}(a), we illustrate another advantage of our approach: it can escape local minima in the optimization landscape that gradient-based methods would get trapped in. There we compare, for recompilation of a single Trotter step, the compilation `energy' $\langle H_{\text{comp}} \rangle$, the covariance vector norm $\|\bm{f}\|^2$ (where $\bm{f}$ is a vector of covariances $\langle O_j O_k \rangle-\langle O_j \rangle \langle O_k \rangle$ of 3-local observables $O_j$), and the step fidelity $\mathcal{F}_{\bm{\theta}}\equiv |\bra{\psi}U^\dagger(\bm{\theta})V(\Delta t)U(\bm{\theta}_{\bm{t}})\ket{\psi}|^2$, where $V(\Delta t)$ is a second-order Trotter step over time $\Delta t$. The CoVaR method directly minimizes $\|\bm{f}\|^2$, which also leads to direct reduction of the infidelity $1-f_{\bm{\theta}}$. However, this includes an \emph{increase} in the compilation energy $\langle H_{\text{comp}} \rangle$, taking a path in parameter space that improves fidelity which gradient-based methods would not take due to the energy increase. We note that at large system sizes, the local observables $O_j$ included in $\bm{f}$ should be chosen randomly at each step, but at small system sizes it remains tractable to use all 3-local observables.

In \cref{fig:covar_trap_escape}(b), we depict the hardware-efficient ansatz used for variational dynamics in \cref{sec:nisq_dos}. We recompile second-order Trotter steps of size $\Delta t = 0.2$ using 25 steps, each corresponding to a change of parameters and new classical shadow. The resulting circuit parameters are then used to compute Loschmidt echoes under the influence of the noise model outlined in \cref{appendix:ibm_eagle_model}.

As noted before, this method can be used for global-phase-insensitive recompilation of dynamics (as studied in Refs. \cite{li2017efficient,cirstoiu2020variational,barison2021efficient,berthusen2022quantum,goh2023lie}) using the circuit structure of \cref{fig:compilation_circuits}(b). For the purposes of this work, we perform our optimization noiselessly on a classical computer, but we anticipate that our approach may be a competitive method of variationally approximating dynamics in general due to the shot-frugal classical shadows approach. We anticipate producing a deeper investigation of this approach in future; further steps in this direction are discussed in Ref.~\cite{goh2024protocols}.

\end{document}